\documentclass[pra,aps,showpacs,onecolumn,twoside,superscriptaddress]{revtex4}

\usepackage{amsmath,amsfonts,amssymb,color,epsfig,graphics,graphicx,latexsym,revsymb,theorem,url,verbatim}

\newtheorem{definition}{Definition}
\newtheorem{proposition}[definition]{Proposition}
\newtheorem{lemma}[definition]{Lemma}

\newtheorem{theorem}[definition]{Theorem}
\newtheorem{corollary}[definition]{Corollary}
\newtheorem{conjecture}[definition]{Conjecture}

\newtheorem{remark}[definition]{Remark}
\newtheorem{example}[definition]{Example}

\def\squareforqed{\hbox{\rlap{$\sqcap$}$\sqcup$}}
\def\qed{\ifmmode\squareforqed\else{\unskip\nobreak\hfil
\penalty50\hskip1em\null\nobreak\hfil\squareforqed
\parfillskip=0pt\finalhyphendemerits=0\endgraf}\fi}
\def\endenv{\ifmmode\;\else{\unskip\nobreak\hfil
\penalty50\hskip1em\null\nobreak\hfil\;
\parfillskip=0pt\finalhyphendemerits=0\endgraf}\fi}
\newenvironment{proof}{\noindent \textbf{{Proof.~} }}{\qed}

\def\bcj{\begin{conjecture}}
\def\ecj{\end{conjecture}}
\def\bcr{\begin{corollary}}
\def\ecr{\end{corollary}}
\def\bd{\begin{definition}}
\def\ed{\end{definition}}
\def\bea{\begin{eqnarray}}
\def\eea{\end{eqnarray}}
\def\bem{\begin{enumerate}}
\def\eem{\end{enumerate}}
\def\bim{\begin{itemize}}
\def\eim{\end{itemize}}
\def\bl{\begin{lemma}}
\def\el{\end{lemma}}
\def\bpf{\begin{proof}}
\def\epf{\end{proof}}
\def\bpp{\begin{proposition}}
\def\epp{\end{proposition}}
\def\br{\begin{remark}}
\def\er{\end{remark}}
\def\bt{\begin{theorem}}
\def\et{\end{theorem}}

\def\a{\alpha}

\def\d{\delta}

\def\ve{\varepsilon}
\def\z{\zeta}

\def\l{\lambda}
\def\m{\mu}

\def\x{\xi}

\def\r{\rho}
\def\s{\sigma}

\def\c{\chi}
\def\ps{\psi}
\def\o{\omega}

\def\G{\Gamma}

\def\S{\Sigma}

\def\GL{{\mbox{\rm GL}}}

\newcommand{\nc}{\newcommand}

\nc{\cA}{{\cal A}} \nc{\cB}{{\cal B}} \nc{\cC}{{\cal C}}
\nc{\cD}{{\cal D}} \nc{\cE}{{\cal E}} \nc{\cF}{{\cal F}}
\nc{\cG}{{\cal G}} \nc{\cH}{{\cal H}} \nc{\cI}{{\cal I}}
\nc{\cJ}{{\cal J}} \nc{\cK}{{\cal K}} \nc{\cL}{{\cal L}}
\nc{\cM}{{\cal M}} \nc{\cN}{{\cal N}} \nc{\cO}{{\cal O}}
\nc{\cP}{{\cal P}} \nc{\cR}{{\cal R}} \nc{\cS}{{\cal S}}
\nc{\cT}{{\cal T}} \nc{\cU}{{\cal U}} \nc{\cV}{{\cal V}}
\nc{\cW}{{\cal W}} \nc{\cX}{{\cal X}} \nc{\cZ}{{\cal Z}}

\nc{\hA}{{\hat{A}}} \nc{\hB}{{\hat{B}}} \nc{\hC}{{\hat{C}}}
\nc{\hD}{{\hat{D}}} \nc{\hE}{{\hat{E}}} \nc{\hF}{{\hat{F}}}
\nc{\hG}{{\hat{G}}} \nc{\hH}{{\hat{H}}} \nc{\hI}{{\hat{I}}}
\nc{\hJ}{{\hat{J}}} \nc{\hK}{{\hat{K}}} \nc{\hL}{{\hat{L}}}
\nc{\hM}{{\hat{M}}} \nc{\hN}{{\hat{N}}} \nc{\hO}{{\hat{O}}}
\nc{\hP}{{\hat{P}}} \nc{\hR}{{\hat{R}}} \nc{\hS}{{\hat{S}}}
\nc{\hT}{{\hat{T}}} \nc{\hU}{{\hat{U}}} \nc{\hV}{{\hat{V}}}
\nc{\hW}{{\hat{W}}} \nc{\hX}{{\hat{X}}} \nc{\hZ}{{\hat{Z}}}

\def\diag{\mathop{\rm diag}}
\def\dim{\mathop{\rm Dim}}
\def\es{\emptyset}

\def\lin{\mathop{\rm span}}

\def\max{\mathop{\rm max}}
\def\min{\mathop{\rm min}}

\def\op{\oplus}

\def\rank{\mathop{\rm rank}}

\def\tr{\mathop{\rm Tr}}

\def\bigox{\bigotimes}
\def\dg{\dagger}

\def\ox{\otimes}
\def\ra{\rightarrow}

\def\se{\subseteq}

\newcommand{\bra}[1]{\langle#1|}
\newcommand{\ket}[1]{|#1\rangle}
\newcommand{\proj}[1]{| #1\rangle\!\langle #1 |}
\newcommand{\ketbra}[2]{|#1\rangle\!\langle#2|}
\newcommand{\braket}[2]{\langle#1|#2\rangle}

\newcommand{\jpa}{J. Phys. A}

\newcommand{\pla}{Phys. Lett. A}

\font\germ=eufm10

\def\gm{{\mbox{\germ m}}}

\def\bC{{\mbox{\bf C}}}

\begin{document}
\title{Properties and construction of extreme bipartite states
having positive partial transpose}

\author{Lin Chen}
\affiliation{Department of Pure Mathematics and Institute for
Quantum Computing, University of Waterloo, Waterloo, Ontario, N2L
3G1, Canada} \affiliation{Centre for Quantum Technologies, National
University of Singapore, 3 Science Drive 2, Singapore 117542}
\email{cqtcl@nus.edu.sg (Corresponding~Author)}

\def\Dbar{\leavevmode\lower.6ex\hbox to 0pt
{\hskip-.23ex\accent"16\hss}D}
\author {{ Dragomir {\v{Z} \Dbar}okovi{\'c}}}

\affiliation{Department of Pure Mathematics and Institute for
Quantum Computing, University of Waterloo, Waterloo, Ontario, N2L 3G1, Canada} \email{djokovic@uwaterloo.ca}

\begin{abstract}
We consider a bipartite quantum system $\cH_A\ox\cH_B$ with
$M=\dim\cH_A$ and $N=\dim\cH_B$. We study the set $\cE$ of extreme
points of the compact convex set of all states having positive
partial transpose (PPT) and its subsets
$\cE_r=\{\r\in\cE:\rank\r=r\}$. Our main results pertain to the
subsets $\cE_r^{M,N}$ of $\cE_r$ consisting of states whose reduced
density operators have ranks $M$ and $N$, respectively. The set
$\cE_1$ is just the set of pure product states. It is known that
$\cE^{M,N}_r=\es$ for $1< r\le\min(M,N)$ and for $r=MN$. We prove
that also $\cE_{MN-1}^{M,N}=\es$. Leinaas, Myrheim and Sollid have
conjectured that $\cE_{M+N-2}^{M,N}\ne\es$ for all $M,N>2$ and that
$\cE_r^{M,N}=\es$ for $1<r<M+N-2$. We prove the first part of their
conjecture. The second part is known to hold when $\min(M,N)=3$ and
we prove that it holds also when $\min(M,N)=4$. This is a
consequence of our result that $\cE_{N+1}^{M,N}=\emptyset$ if
$M,N>3$.

We introduce the notion of ``good'' states, show that all pure
states are good and give a simple description of the good
separable states. For a good state $\r\in\cE_{M+N-2}^{M,N}$, we
prove that the range of $\r$ contains no product vectors and
that the partial transpose of $\r$ has rank $M+N-2$ as well. In the special case $M=3$, we construct good $3\times N$ extreme states of rank $N+1$ for all $N\ge4$.

\end{abstract}

\date{ \today }




\maketitle

\tableofcontents


%
%
%
%
%
%

\section{\label{sec:introduction} Introduction}

Let us consider a finite-dimensional bipartite quantum systems
represented by the complex Hilbert space $\cH=\cH_A\ox\cH_B$
with $\dim\cH_A=M$ and $\dim\cH_B=N$. A {\em state} of this
system is a positive semidefinite linear operator $\r:\cH\to\cH$
with $\tr\r=1$. A {\em pure state} is a state $\r=\proj{\psi}$ where
$\ket{\psi}\in\cH$ is a unit vector.
A {\em product state} is a state $\r=\r_1\ox\r_2$ where $\r_1$
is a state on $\cH_A$ and $\r_2$ a state on $\cH_B$.
If moreover $\r_1$ and $\r_2$ are pure states, then we say that
$\r=\r_1\ox\r_2$ is a {\em pure product state}.
For any nonzero vector $\ket{x,y}:=\ket{x}\ox\ket{y}$ we say
that it is a {\em product vector}.
By definition, a {\em separable state}, say $\s$, is a convex linear
combination of pure product states $\r_i$:
 \bea \label{ConvexComb}
\s=\sum_{i=1}^k p_i\r_i, \quad p_i\ge0,\quad \sum_{i=1}^k p_i=1.
 \eea
A state is {\em entangled} if it is not separable. It is a highly
nontrivial task to determine whether a given bipartite state is
separable \cite{gurvits03}.

We can write any linear operator $\r$ on $\cH$ as
 \bea \label{OperatorRo}
\r=\sum_{i,j=0}^{M-1} \ketbra{i}{j}\ox\r_{ij},
 \eea
where $\{\ket{i}:0\le i<M\}$ is an orthonormal (o.n.) basis of $\cH_A$
and $\r_{ij}=\bra{i}\r\ket{j}$ are linear operators on $\cH_B$.
Then the partial transpose $\r^\G$ of $\r$ is defined by
 \bea \label{ParcijTransp}
\r^\G=\sum_{i,j=0}^{M-1} \ketbra{j}{i}\ox\r_{ij}.
 \eea
The {\em reduced operators} $\r_A$ and $\r_B$ of $\r$ are defined by
 \bea \label{ParcijTragA}
\r_A={\rm Tr}_B(\r)=\sum_{i,j=0}^{M-1}
\tr(\r_{ij}) \ketbra{i}{j},\quad
\r_B={\rm Tr}_A(\r)=\sum_{i=0}^{M-1} \r_{ii},
\eea
where $\tr_A$ and $\tr_B$ are partial traces.
We refer to $\rank\r_A$ as the {\em A-local rank} and to
$\rank\r_B$ as the {\em B-local rank} of $\r$. We shall use the
following very convenient but non-standard terminology.
 \bd \label{df:kxlstate} A bipartite state $\r$ is a {\em $k\times l$
state} if $\rank\r_A=k$ and $\rank\r_B=l$. \ed

If $\r$ is a separable state, then necessarily $\r^\G\ge0$ (i.e.,
$\r^\G$ is positive semidefinite). This necessary condition for
separability is due to Peres \cite{peres96}. If $\dim\cH\le6$ then
this separability condition is also sufficient (but not otherwise)
\cite{hhh96,horodecki97,choi82,woronowicz76}. We say that a state
$\r$ is PPT if it satisfies the Peres condition $\r^\G\ge0$. A state
$\r$ is NPT if $\r^\G$ is not positive semidefinite.

We are interested in the problem of describing the set, $\cE$,
of extreme points of the compact convex set of all PPT states.
We shall refer to any $\r\in\cE$ as an {\em extreme state}.
Since every PPT state is a convex linear combination of extreme
states, it is important to understand the structure of $\cE$.
The rank function provides the partition
 \bea \label{Partition-1}
\cE=\bigcup_{r=1}^{MN} \cE_r,\quad \cE_r:=\{\r\in\cE:\rank\r=r\}.
 \eea
The first part, $\cE_1$, is the set of pure product states. Let us briefly explain this observation. Let $\r\in\cE_1$. Since any state of rank one is pure, $\r$ is a pure PPT state. It follows from the Schmidt decomposition that any pure PPT state is necessarily a product state. Thus $\r$ is a pure product state.
Conversely, let $\r$ be a pure product state. As $\r$ has rank
one, it is extremal among all states, and in particular it is
extremal among all PPT states. Thus $\r\in\cE_1$.

It follows easily that $\cE_1$ is also the set of extreme points of the compact convex set consisting of all separable states.
Consequently, for $r>1$, the set $\cE_r$ contains only entangled states. Since all PPT states of rank less than four are separable
\cite{hst99,hlv00}, we have $\cE_2=\cE_3=\emptyset$ (see also  Proposition \ref{prop:PPTMxNrankN} below).

We can further partition the subsets $\cE_r$ by using the local ranks
 \bea \label{Partition-2}
\cE_r=\bigcup_{(k,l)} \cE_r^{k,l},\quad
\cE_r^{k,l}:=\{\r\in\cE_r:\rank\r_A=k,\;\rank\r_B=l\}.
 \eea
For $r=1$ we have $\cE_1=\cE_1^{1,1}$ and for $r=4$ we have
$\cE_4=\cE_4^{3,3}$.

Assume that $r>4$. Then the problem of deciding which sets
$\cE_r^{k,l}$ are nonempty is apparently hard. If $kl\le6$ then any
$k\times l$ PPT state is separable, and so $\cE_r^{k,l}=\emptyset$.
It is easy to see that the condition $\min(k,l)>1$ is necessary for
$\cE_r^{k,l}$ to be nonempty. The condition $\max(k,l)<r$ is also
necessary. This follows from the well known facts that any $k\times
l$ state of rank $r$ is NPT if $r<\max(k,l)$ and is separable if
$r=\max(k,l)$. See Theorem \ref{thm:PPTMxNrank<M,N} and Proposition
\ref{prop:PPTMxNrankN} which are proved in \cite{hst99} and
\cite{kck00}, respectively.

As we shall see below, $\cE_5^{4,4}=\emptyset$, and so for $r=5$
we have
 \bea \label{Partition-3}
\cE_5=\cE_5^{2,4}\cup\cE_5^{4,2}\cup\cE_5^{3,3}\cup\cE_5^{3,4}\cup \cE_5^{4,3}.
 \eea
It was shown in \cite{agk10} that the sets $\cE_5^{2,4}$ and
$\cE_5^{4,2}$ are nonempty.
The sets $\cE_5^{3,4}$ and $\cE_5^{4,3}$ are also nonempty,
see Example \ref{Kon-Mnogo}. Since the example of the $3\times3$
PPTES of rank five constructed in \cite{clarisse06} is extreme,
it follows that $\cE_5^{3,3}\ne\emptyset$.

The following conjecture was proposed recently by Leinaas,
Myrheim and Sollid \cite[Sec. III, part D]{lms10}.
 \bcj
 \label{conj:LMS2010highdimension}
$(M,N>2)$

(i) $\cE_{M+N-2}^{M,N}\ne\emptyset$;

(ii) $\cE_r^{M,N}=\emptyset$ for $1<r<M+N-2$;

(iii) if $\r\in\cE_{M+N-2}^{M,N}$ then $\rank(\r^\G)=M+N-2$;
 \ecj

Note that part (ii) of this conjecture is true if $\min(M,N)=3$.
It is also true if $\min(M,N)=4$.
This is a consequence of the general fact,
proved in Theorem \ref{thm:MxNrank(N+1)=reducible,M>3},
which says that $\cE_{N+1}^{M,N}=\emptyset$ if $M,N>3$.
(In particular, we have $\cE_5^{4,4}=\emptyset$.)
Consequently, if $k,l>3$ then the condition $\max(k,l)<r-1$ is
necessary for $\cE_r^{k,l}$ to be nonempty.
The question whether (ii) is valid when $\min(M,N)>4$ remains open.

In Theorem \ref{thm:MxNrank(M+N-2)} we prove part (i) of
the above conjecture.
The proof is based on explicit construction of the required extreme
states. An important tool used in this proof is the extremality
criterion first discovered in \cite{lmo07}, and independently in
\cite{agk10}
(see Proposition \ref{prop:ExtCrit} for an enhanced version).
It has been hard in the past to verify that a given PPT state is
extreme, see e.g. \cite{ha09,kk07}.
By using the extremality criterion, this is now a routine task.
Proposition \ref{prop:extremeNecCond} gives a simple necessary
condition for extremality. The well known fact that
$\cE_{MN}^{M,N}=\emptyset$ is an immediate consequence of this
proposition. By using the same proposition, we prove that also
$\cE_{MN-1}^{M,N}=\emptyset$, see Corollary \ref{cor:upperbound}.

Extreme states have applications to some important problems of
entanglement theory. First, it is known that extreme states of rank $>1$ are also edge states \cite{agk10}.
We recall that a PPT state $\r$ is an {\em edge state} if there is no product vector $\ket{a,b}\in\cR(\r)$ such that $\ket{a^*,b}\in\cR(\r^\G)$. Note that any edge state is necessarily entangled. Second, entanglement
distillation is a core task in quantum information theory
\cite{bds96}. Although not all entangled states can be distilled
\cite{hhh98}, we will show that extreme states can play the role of activators in entanglement distillation in Sec.
\ref{sec:application}. Third, characterizing extreme states is
useful for solving the separability problem in some special cases, see Proposition \ref{pp:TeoremaSep} (ii).

After examining many examples of bipartite states, we came to the
conclusion that they should be divided into two broad categories.
For lack of a better name, we refer to them as ``good'' and ``bad''
states. The characterization of these states, in particular good
states is the main problem of this paper. More generally, we shall
first define these notions for vector subspaces of $\cH$. For this
we shall make use of complex projective spaces and some basic facts
from algebraic geometry. We shall recall these notions and facts in
the next section. For more information the reader may consult
\cite{h92}.

We denote by $\cP_{AB}$ the complex projective space of (complex)
dimension $MN-1$ associated to $\cH$, and denote by $\cP_A$ and
$\cP_B$ the projective spaces associated to $\cH_A$ and $\cH_B$,
respectively. For any vector subspace $K\subseteq\cH$, we denote by
$\hat{K}$ the projective subspace of $\cP_{AB}$ associated to $K$.
By a {\em projective variety} we mean any Zariski closed subset, say
$X$, of $\cP_{AB}$. If $X$ is not the union of two proper Zariski
closed subsets, then we say that $X$ is {\em irreducible}. Any
projective variety $X$ is a finite irredundant union of irreducible
projective varieties $X_i$, $i=1,\ldots,s$. The $X_i$ are unique up
to indexing, and we refer to them as the {\em irreducible
components} of $X$.

The points of $\cP_{AB}$ which correspond to product vectors form
a projective variety $\S=\S_{M-1,N-1}$ known as the {\em Segre
variety}. Thus a point of $\S$ is a 1-dimensional subspace spanned by a product vector. The variety $\S$ is isomorphic to
the direct product $\cP_A\times\cP_B$.
Its complex dimension is $M+N-2$, and its codimension in the
ambient projective space $\cP_{AB}$ is $(M-1)(N-1)$.
Let us briefly recall the concept of degree for a projective
variety, say $X$, of dimension $k$ embedded in the projective space $\cP^n$. The {\em degree} of $X$ can be defined as the
number of intersection points of $X$ with a general projective
subspace $L$ of complementary dimension, $n-k$. For instance, for
the Segre variety $\S\subset\cP_{AB}$, we have to take $L$ of
dimension $(MN-1)-(M+N-2)=(M-1)(N-1)$. We also recall (see
\cite[Example 18.15]{h92}) that the degree of $\S$ is the
binomial coefficient
\begin{equation} \label{jed:delta}
 \d=\d(M,N) :=\binom{M+N-2}{M-1}.
\end{equation}

 \bd \label{def:Good Subspace}
Let $K\subseteq\cH$ be a vector subspace of dimension $k+1$ and
let $X=\hat{K}\cap\S$. We say that $K$ is {\em good} in $\cH$ if either
$X=\emptyset$ or $\dim X=k-(M-1)(N-1)\ge0$ and $\hat{K}$ and $\S$
intersect generically transversely.
Otherwise, we say that $K$ is {\em bad} in $\cH$.
 \ed

Let us clarify this definition. If $k<(M-1)(N-1)$ then there
are two possibilities: $X=\es$ and so $K$ is good or
$X\ne\es$ and so $K$ is bad.
Assume now that $k\ge(M-1)(N-1)$. Then necessarily $X\ne\es$
(see Proposition \ref{prop:Intersection} below), and let
$X_i$, $i=1,\ldots,s$, be the irreducible components of $X$.
Moreover, we have $\dim X_i\ge k-(M-1)(N-1)$ for each index $i$.
As $\dim X=\max_i\dim X_i$, the assertion $\dim X=k-(M-1)(N-1)$
is equivalent to the assertion
that $\dim X_i=k-(M-1)(N-1)$ for each index $i$. Finally, the
transversality condition means that for each $i$ there exists a
point $x_i\in X_i$ such that $x_i\notin X_j$ for $j\ne i$ and
the sum of the tangent spaces of $\hat{K}$ and $\S$ at $x_i$ is
equal to the whole tangent space of $\cP_{AB}$ at the same
point. To state an
affine equivalent of this condition, for any product vector
$\ket{a,b}$, we set $S_{a,b}=\ket{a}\ox\cH_B+\cH_A\ox\ket{b}$.
If the point $x_i$ is represented by the product vector
$\ket{a_i,b_i}$, then the transversality condition at $x_i$ is
equivalent to the equality $K+S_{a_i,b_i}=\cH$.

To any state $\r$, we attach a (possibly empty) projective
variety $X_\r:=\hat{K}\cap\S$ where $K=\ker\r$. We say that $\r$ is {\em good} or {\em bad} in $\cH$ if $K$ is good or bad in $\cH$, respectively.
In Theorem \ref{thm:goodseparablestate} we give a simple description of good separable states.
In Theorem \ref{thm:delta} we prove that
if $\r$ is a PPT state of rank $r$ and $\ker\r$ contains no
2-dimensional subspace $V\ox W$, then either $r=M+N-2$ and
$|X_\r|=\d$ or $r>M+N-2$ and $|X_\r|<\d$.
(By $|X|$ we denote the cardinality of a set $X$.)

The good states have many good properties. First note that if $\r$
is a good state and $\s$ another state having the same kernel (or,
equivalently, range) as $\r$, then $\s$ is also good.
Second, if
$\r$ is a good state then the same is true for its transform
$\r'=V\r V^\dag$ by any invertible local operator (ILO) $V=A\ox B$.
Since $\ker\r=V^\dag\ker\r'$, this follows from the fact that
$V^\dag=A^\dag\ox B^\dag$ acts on $\cP_{AB}$ as a projective
transformation and maps $\S$ onto itself. By using these two facts,
we can construct new good states and study their properties. For
instance, we prove in Theorem \ref{thm:goodnondistillable} that good
entangled non-distillable states must have full local ranks. One of
the most important conjectures in entanglement theory asserts that
some of the Werner states \cite{werner89}, which have full rank and
thus are good, are not distillable \cite{dss00}. On the other hand,
we show in Proposition \ref{prop:puregood} that all pure states (the
basic ingredients for quantum-information tasks) are good. Third, in
Theorem \ref{thm:good,PPT}, we provide a family of good separable
states, which turn out to be the generalized classical states which
occur in the study of quantum discord \cite{ccm11}. Hence, these
examples show the operational meaning of good states in quantum
information.

Part (ii) of Theorem \ref{thm:LMS2010(ii),ker=finitePRO} states that if $\r$ is a good $M\times N$ PPT state of rank $M+N-2$, then the same holds true for $\r^\G$. Consequently, part (iii) of
Conjecture \ref{conj:LMS2010highdimension} is valid in the good case, while it remains open in the bad case.
We say that a subspace of $\cH$ is a {\em completely entangled
subspace} (CES) if it contains no product vectors. It follows from part (iii) of Theorem \ref{thm:LMS2010(ii),ker=finitePRO}
that if $\r\in\cE_{M+N-2}^{M,N}$ is good and $M,N>2$, then
$\cR(\r)$ and $\cR(\r^\G)$ are CES.

In the borderline case, $r=M+N-2$, we shall propose another
conjecture. Let us first introduce the following definition.
 \bd \label{def:StronglyExtreme}
A PPT state $\s$ is {\em strongly extreme} if there are no PPT states $\r\ne\s$ such that $\cR(\r)=\cR(\s)$.
 \ed
Obviously any pure product state is strongly extreme.
It follows from Proposition \ref{prop:3x3rank4PPTES,samerange} that
any $3\times3$ PPTES of rank four is also strongly extreme.
The strongly extreme states are extreme,
see Lemma \ref{le:stronglyextreme}. In the same lemma it is
shown that the range of a strongly extreme state is a CES.
There exist examples of extreme states which are not strongly extreme,
e.g., $3\times3$ extreme states $\s$ of rank five or six,
see \cite{kkl11} and its references.
Indeed, since $\rank\s\ge5$, $\cR(\s)$ is not a CES (see
Proposition \ref{prop:Intersection}).

We can now state our conjecture which generalizes
Proposition \ref{prop:3x3rank4PPTES,samerange}.
 \bcj \label{conj:EkstrSt}
Every state $\r\in\cE_{M+N-2}^{M,N}$, $M,N>2$, is strongly extreme.
 \ecj
Theorem \ref{thm:LMS2010(ii),ker=finitePRO} also shows that
Conjecture \ref{conj:EkstrSt} is valid in the good case,
but it remains open in the bad case.

The content of our paper is as follows.

Sec. \ref{sec:preliminary} has two subsections.
In the first one we describe the tools that enable us to
represent bipartite density matrices and perform the basic local
operations on them. We also introduce the necessary background
and give references about complex projective varieties embedded
in an ambient complex projective space, $\cP_{AB}$ in our context.
We also define the good and bad subspaces and states.
In the second subsection we summarize some important facts from
quantum information theory that we will need. We introduce the
concept of reducible and irreducible bipartite states, present
the extremality criterion and give a short proof.

Sec. \ref{sec:goodsep} deals mostly with the properties of good
states. We first single out a special class of good states which
we call universally good. These are good states which remain
good after embedding the original $M\ox N$ quantum system into
an arbitrary $M'\ox N'$ quantum system with $M'\ge M$ and
$N'\ge N$.
Then we show that all pure states are good, and consequently they are also universally good (see Proposition \ref{prop:puregood}).
The universally good PPT states are fully characterized in
Theorem \ref{thm:good,PPT}, in particular they are separable.
The Proposition \ref{prop:MxN,>delta=infinitelymany} relates the number, $m$, of product vectors in a subspace $H\subseteq\cH$
to the dimension of $H$. In particular, it is shown that if $m=\d$
then this dimension must be $(M-1)(N-1)+1$. In Theorem
\ref{thm:irreducible=segre} we show how one can find all
irreducible components of the variety $X_\r$ for arbitrary separable
state $\r$. It turns out each of these components is the Segre
variety of a subspace $V\ox W\se\ker\r$. Finally, we obtain a very
simple characterization of the good separable states in Theorem
\ref{thm:goodseparablestate}.

Sec. \ref{sec:generic} is mainly about the borderline case: the
$M\times N$ PPT states of rank $M+N-2$. We need two results from
algebraic geometry, which are proved in the Appendix. There are two
subsections. In the first one we prove a general result which
applies to all $M\times N$ PPT states $\r$, namely Theorem
\ref{thm:delta}. First, it shows that if $X_\r$ is an infinite set
then $\ker\r$ contains a 2-dimensional subspace $V\ox W$. (For a
stronger version of this result see Theorem
\ref{thm:(M+N-r-1)x1=1x(M+N-r-1)ppt}). Second, if $m:=|X_\r|<\infty$
then either $m=\d$ and $\rank\r=M+N-2$ or $m<\d$ and
$\rank\r>M+N-2$. By using Theorem \ref{thm:delta}, we prove in
Corollary \ref{cor:upperbound} that $\cE_{MN-1}^{M,N}=\es$. In the
second subsection we characterize good $M\times N$ PPT states of
rank $M+N-2$, see Theorem \ref{thm:LMS2010(ii),ker=finitePRO}.
Proposition \ref{prop:NECESSARY=good} shows that if $\r$ is an
$M\times N$ state, $|X_\r|<\infty$ and $\rank\r^\G=M+N-2$, then $\r$
must be a good PPT state of the same rank.

In Sec. \ref{sec:N+1} we investigate the $M\times N$ PPT states $\r$
of rank $N+1$. In Proposition \ref{prop:MxNrank(N+1)PPT} we
characterize such states $\r$ when the range of $\r$ contains at
least one product vector. In Theorem \ref{thm:MxNrank(N+1)reducible}
we analyze further the case when $\r$ is entangled. The main result
of this section is that $\r$ cannot be extreme when $M,N>3$, see
Theorem \ref{thm:MxNrank(N+1)=reducible,M>3}. Based on this result
we construct a link between the good and extreme states in
Proposition \ref{pp:TeoremaSep}. Then in Theorem
\ref{thm:MxNrank(N+1)=3exclusive} we extend assertions (i-ii) of
Theorem \ref{thm:3x3PPTstates} to $M\ox N$ systems. We also give a
sufficient condition for extremality of $3\times N$ states of rank
$N+1$, see Theorem \ref{thm:3xN-PPTESbirank}.

In Sec. \ref{sec:example}, we construct many examples of good and
bad $M\times N$ PPT states of rank $M+N-2$.
There are two subsections; the first contains good cases and
the second bad cases.
The most important are the infinite families given in Examples
\ref{Besk-Fam,KERNEL=finite,M=3,Nge4} and \ref{eg:Mge3,Nge4,extreme}.
The first of these families consists of strongly extreme $3\times N$
states of rank $N+1$, $N>3$, and we prove that all of these states
are good (see Theorem \ref{thm:PrvaFam}).
The second family consists of bad $M\times N$ PPT states of rank
$M+N-2$. In Theorem \ref{thm:MxNrank(M+N-2)}, we prove that all
of these states are extreme.
Thus, we confirm part (i) of
Conjecture \ref{conj:LMS2010highdimension}.

In Sec. \ref{sec:application} we propose some open problems.

\section{\label{sec:preliminary}
Preliminaries}

In this section we state our conventions and notation,
and review known and derive some new results which will be
used throughout the paper.

We shall write $I_k$ for the identity $k\times k$ matrix.
We denote by $\cR(\r)$ and $\ker \r$ the range
and kernel of a linear map $\r$, respectively. Many of the results
will begin with a clause specifying the assumptions on $M$ and $N$.
The default will be that $M,N>1$. From now on, unless stated
otherwise, the states will not be normalized.

We say that a non-normalized state $\r$ is {\em extreme} if its
normalization is an extreme point of the set of normalized
PPT states.
Equivalently, a non-normalized state $\r$ is extreme if it is PPT
and cannot be written as the sum of two non-proportional PPT states.

The rest of this section is divided into two parts. The first part
deals with mathematical topics and the second one with quantum
information.

\subsection{Mathematics}

We shall denote by $\{\ket{i}_A:i=0,\ldots,M-1\}$ and
$\{\ket{j}_B:j=0,\ldots,N-1\}$ o.n. bases of $\cH_A$ and $\cH_B$,
respectively. The subscripts A and B will be often omitted. Any
state $\r$ of rank $r$ can be represented as
(see \cite[Proposition 6]{cd11JPA})
 \bea \label{MxN-State}
\r=\sum_{i,j=0}^{M-1} \ketbra{i}{j}\ox C_i^\dag C_j,
 \eea
where the $C_i$ are $R\times N$ matrices and $R$ is an arbitrary
integer $\ge r$. In particular, one can take $R=r$. We shall often
consider $\r$ as a block matrix $\r=C^\dag C=[C_i^\dag C_j]$, where
$C=[C_0~C_1~\cdots~C_{M-1}]$ is an $R\times MN$ matrix.
Thus $C_i^\dag C_j$ is the matrix of the linear operator
$\bra{i}_A \r \ket{j}_A$ acting on $\cH_B$.
For the reduced density matrices, we have the formulae
 \bea \label{RedStates}
\r_B=\sum_{i=0}^{M-1} C_i^\dag C_i; \quad
\r_A=[\tr C_i^\dag C_j], \quad i,j=0,\ldots,M-1.
 \eea

It is easy to verify that the range
of $\r$ is the column space of the matrix $C^\dag$ and that
 \bea \label{JezgroRo}
\ker\r=\left\{\sum_{i=0}^{M-1}\ket{i}\ox\ket{y_i}:
\sum_{i=0}^{M-1}C_i\ket{y_i}=0\right\}.
 \eea
In particular, if $C_i\ket{j}=0$ for some $i$ and $j$ then
$\ket{i,j}\in\ker\r$.

For any bipartite state $\r$ we have
 \bea
 \label{ea:rhoBGamma=rhoB}
\left( \r^\G \right)_B &=& {\tr}_A \left( \r^\G \right) =
{\tr}_A \r = \r_B, \\
 \label{ea:rhoAGamma=rhoGammaA}
\left( \r^\G \right)_A &=& {\tr}_B \left( \r^\G \right) = \left(
{\tr}_B \r \right)^T = ( \r_A )^T. \eea
(The exponent T denotes transposition.)
Consequently,
 \bea
 \rank \left( \r^\G \right)_A = \rank \r_A,
 \quad
 \rank \left( \r^\G \right)_B = \rank \r_B.
 \eea
If $\r$ is an $M\times N$ PPT state, then $\r^\G$ is too. If $\r$ is
a PPTES so is $\r^\G$, but they may have different ranks.
We refer to the ordered pair $(\rank\r,\rank\r^\G)$ as the {\em
birank} of $\r$.

For counting purposes, we do not distinguish two product vectors
which are scalar multiples of each other. The maximum dimension of a
CES is $(M-1)(N-1)$, see \cite{parth04,ws08}. For an explicit and
simple construction of CES with this dimension see
\cite{parth04,bhat06}. For convenience, we shall represent the pure
state $\sum_{i,j} \x_{ij}\ket{i}\ox\ket{j}$ also by the $M\times N$
matrix $[\x_{ij}]$. Then pure product states are represented by
matrices of rank one.

The {\em partial conjugate} of a product vector $\ket{a,b}$
is the product vector $\ket{a^*,b}:=\ket{a^*}\ox\ket{b}$, where
$\ket{a^*}$ is the conjugate of the vector $\ket{a}$ computed in the
basis $\{\ket{i}_A:i=0,\ldots,M-1\}$. Since
$\ket{a,b}=\ket{za,z^{-1}b}$ for any nonzero $z\in\bC$, and the
partial conjugate of $\ket{za,z^{-1}b}$ is
$\ket{z^*a^*,z^{-1}b}=(z^*/z)\ket{a^*,b}$, we see that the partial
conjugation operation on product vectors is well-defined only up to
a phase factor.
Thus, the partial conjugation is an involutory automorphism of $\S$
viewed as a real (but not complex) manifold.

In some of the proofs we shall use some basic facts about the
intersection of two projective varieties embedded in a bigger
projective space. Let us briefly describe these facts.
The {\em degree} of a projective variety, say $X$, of dimension $k$
embedded in the projective space $\cP^n$ can be defined as the
number of intersection points of $X$ with a general projective
subspace $L$ of complementary dimension, $n-k$. For instance, for
the Segre variety $\S\subset\cP_{AB}$, we have to take $L$ of
dimension $(MN-1)-(M+N-2)=(M-1)(N-1)$. Recall that the degree of
$\S$ is given by the formula (\ref{jed:delta}),
while every projective subspace has degree 1.

The following proposition is a special case of some basic and
well-known facts from algebraic geometry, see e.g.
\cite[Theorem 7.2]{RH}.
 \bpp
\label{prop:Intersection}
For any projective subspace $L\se\cP_{AB}$ of dimension
$k\ge(M-1)(N-1)$, the intersection $X=L\cap\S$ is nonempty.
Equivalently, any vector subspace of $\cH$ of dimension
$>(M-1)(N-1)$ must contain at least one product vector.
More precisely, if $X_i$ $(i=1,\ldots,s)$ are the irreducible
components of $X$, then
\begin{equation} \label{Presek-Proj}
 \dim X_i\ge k-(M-1)(N-1),\quad i=1,\ldots,s.
\end{equation}
In particular, any vector subspace of $\cH$ of dimension
$>(M-1)(N-1)+1$ contains infinitely many product vectors.
 \epp

If a state $\r$ has kernel of dimension $(M-1)(N-1)+1$
then $\rank\r=M+N-2$. This partially motivates our interest in
states $\r$ of rank $M+N-2$: their kernels must contain at least one
product vector.

Let $\r$ be an $M\times N$ state of rank $r\le M+N-2$ and let
$r'=M+N-1-r$ and $K=\ker\r$.
Denote by $X_i$, $i=1,\ldots,s$, the irreducible components of
$X_\r=\hat{K}\cap\S$ and let $d_i$ be the degree of $X_i$.
If $r\le M+N-2$ then, by Proposition \ref{prop:Intersection},
$\dim X_i\ge r'-1$ for each $i$.

In order to be able to apply the version of the
B\'{e}zout's theorem as stated in
\cite[Bezout's Theorem, pp. 80-81]{mfd76},
we need two conditions: (a) $\dim X_\r=r'-1$ and
(b) $\hat{K}$ and $\S$ intersect generically transversely.
When the conditions (a) and (b) hold, then the B\'{e}zout's theorem
asserts that the following {\em degree formula} is valid
 \bea \label{ea:DegreeFormula}
\delta=\sum_{i=1}^s d_i.
 \eea
At the end of Sec. \ref{sec:goodsep}, we shall verify that good
separable states indeed satisfy this equation.

The more general version of the B\'{e}zout's theorem
\cite[Theorem 18.4]{h92} can be applied
assuming only condition (a). Then the degree formula has to
be replaced by the more general one
 \bea \label{ea:GenDegreeFormula}
\delta=\sum_{i=1}^s \m_id_i,
 \eea
where $\m_i~(\ge1)$ is the intersection multiplicity of $L$ and
$\S$ along $X_i$. The condition (b) implies that all $\m_i=1$
and so (\ref{ea:GenDegreeFormula}) reduces to
(\ref{ea:DegreeFormula}).
The converse is also valid, i.e., if (a) holds then (b) is
equivalent to the assertion that all $\m_i=1$,
see \cite[p. 79]{ps67} and \cite[p. 137-138]{wf98}.

Note that $X_\r=\emptyset$ implies that $r'\le0$, i.e.,
$r>M+N-2$. For the case $r=M+N-2$, we have the following
simple test.

 \bpp
\label{prop:DobroStanje}
If $\r$ is an $M\times N$ (PPT or NPT) state of rank $r=M+N-2$,
then $\r$ is good if and only if $|X_\r|=\d$.
 \epp
 \bpf
Let $K=\ker\r$. In view of the hypotheses, $\r$ is good if and only if $\hat{K}$ and $\Sigma$ intersect transversely. We can apply Eq. \eqref{ea:GenDegreeFormula}. Since the irreducible
components $X_i$ of $X_\r$ are just points, each $d_i=1$ and $s=|X_\r|$. It follows that the equality $|X_\r|=\d$ holds if
and only if each $\mu_i=1$, i.e., if and only if $\hat{K}$ and
$\Sigma$ intersect transversely.
 \epf

As a basic example, we consider the $2\times2$ separable state
$\r=\proj{00}+\proj{11}$. We claim that $\r$ is good. We have to
prove that $K:=\ker\r=\lin\{\ket{01},\ket{10}\}$ is a good
subspace. The variety $X_\r$ consists of only two points,
namely the points corresponding to product vectors
$\ket{a_1,b_1}=\ket{01}$ and $\ket{a_2,b_2}=\ket{10}$. It is easy to verify that $K+S_{a_1,b_1}=K+S_{a_2,b_2}=\cH$. Hence,
the transversality condition is satisfied, and so $\r$ is good.

Another basic example is the $1\times2$ separable state
$\r=\proj{00}+\proj{01}$ in the $2\ox2$ space.
This is a bad state because $\ker\r=\ket{1}\otimes\cH_B$,
and so $\dim X_\r=1$.

Let us mention more examples of good states that are well known
in quantum information.
An o.n. set of product vectors
$\{\psi\}:=\{\ket{\psi_i}:i=1,\ldots,k\}\subset\cH$ is an {\em
unextendible product basis} (UPB) \cite{bdm99} if the subspace
$\{\psi\}^\perp$ is a CES.
If $\{\psi\}$ is a UPB, then the orthogonal projector onto
$\{\psi\}^\perp$ is a PPTES.
It is known that any two-qutrit PPTES $\r$ of rank four can be
constructed by using the fact that there are exactly six product
vectors in $\ker\r$ \cite{cd11JMP}.
Any five of these six product vectors can be converted to an UPB
\cite{bdm99,cd11JMP}.
These $\r$ are good PPTES of the simplest kind.
The UPB construction of PPTES works also in higher dimensions
but is no longer universal.
Indeed, we construct in Example \ref{Besk-Fam,KERNEL=finite} a good
$3\times4$ PPTES $\s$ of rank five with $|X_\s|=\d(3,4)=10$.
Thus $\ker\s$ contains exactly ten product vectors.
Moreover any seven of them are linearly independent.
However, Lemma \ref{le:UPBgeneralposition} shows that no seven of
these ten product vectors can be simultaneously converted by
an ILO to scalar multiples of vectors of an UPB.

In the case $r>M+N-2$ there exist good as well as bad $M\times N$
PPT states of rank $r$.
As examples of good states, we mention the $3\times3$
edge PPTES of birank $(7,6),(7,5)$ and $(5,8)$ constructed in
\cite[Eqs. 5, 6]{hk05}, as well as the one of birank $(6,8)$
constructed very recently \cite[Eq. 1]{ko12}. One can check that the
kernels of these states are CES, and so they are good by the above
definition. As an example of a bad state, we mention the $3\times3$
edge state of rank five constructed in \cite[Sec. II]{clarisse06}.
Its kernel has dimension four and contains exactly two product
vectors. (It is known and easy to check that this state is extreme.)

To avoid possible confusion we give a formal definition of the term
``general position''.

 \bd \label{def:GenPos} We say that a family of product vectors
$\{\ket{\psi_i}=\ket{\phi_i}\ox\ket{\chi_i}:i\in I\}$ is in {\em
general position} (in $\cH$) if for any $J\se I$ with $|J|\le M$
the vectors $\ket{\phi_j}$, $j\in J$, are linearly independent
and for any $K\se I$ with $|K|\le N$ the vectors $\ket{\chi_k}$,
$k\in K$, are linearly independent.
 \ed

We warn the reader that it is possible for a family of product
vectors contained in a subspace $V\ox W\se\cH$ to be in general
position in $V\ox W$ but not in general position in $\cH$.

\subsection{Quantum information}

Let us now recall some basic results from quantum information, for
proving the separability, distillability and PPT properties of some
bipartite states.

We say that two $n$-partite states $\r$ and $\s$ are {\em equivalent
under stochastic local operations and classical communications} (or
{\em SLOCC-equivalent}) if there exists an
ILO $A=\bigox^n_{i=1} A_i$ such that $\r=A\s A^\dg$ \cite{dvc00}.
They are {\em LU-equivalent} if the $A_i$ can be chosen to be unitary.
In most cases of the present work, we will have $n=2$.
It is easy to see that any ILO transforms PPT, entangled, or separable state into the same kind of states.
We shall often use ILOs to simplify the density matrices of states.

From \cite[Theorem 1]{hst99} we have
 \bt
\label{thm:PPTMxNrank<M,N} The $M\times N$ states of rank less than
$M$ or $N$ are distillable, and consequently they are NPT.
 \et

The next result follows from \cite{hlv00} and Theorem
\ref{thm:PPTMxNrank<M,N}, see also \cite[Proposition 6
(ii)]{cd11JPA}.

 \bpp
\label{prop:PPTMxNrankN} If $\r$ is an $M\times N$ PPT state of rank
$N$, then $\r$ is a sum of $N$ pure product states.
Consequently, the rank of any PPTES is bigger than any of its
local ranks, and any PPT state of rank at most three is separable.
 \epp

(By Theorem \ref{thm:PPTMxNrank<M,N}, the hypothesis of this
proposition implies that $M\le N$.)

Let us recall from \cite[Theorem 22]{cd11JPA} and \cite[Theorems
17,22]{cd11JMP} the main facts about the $3\times3$ PPT states of
rank four. Let $M=N=3$ and let $\cU$ denote the set of UPBs in
$\cH=\cH_A\ox\cH_B$. For $\{\psi\}\in\cU$ we denote by $\Pi\{\psi\}$
the normalized state $(1/4)P$, where $P$ is the orthogonal projector
onto $\{\psi\}^\perp$.

\bt \label{thm:3x3PPTstates} $(M=N=3)$ For a $3\times3$ PPT state
$\r$ of rank four, the following assertions hold.

(i) $\r$ is entangled if and only if $\cR(\r)$ is a CES.

(ii) If $\r$ is separable, then it is either the sum of four pure
product states or the sum of a pure product state and a $2\times2$
separable state of rank three.

(iii) If $\r$ is entangled, then

\quad (a) $\r$ is extreme;

\quad (b) $\rank \r^\G=4$;

\quad (c) $\r= A\ox B~ \Pi\{\psi\}~ A^\dag\ox B^\dag$ for some
$A,B\in\GL_3$ and some $\{\psi\}\in\cU$;

\quad (d) $\ker\r$ contains exactly 6 product vectors,
and these vectors are in general position.
\et

In Sec. \ref{sec:generic}, we shall generalize the results (i) and
(ii) to arbitrary bipartite systems. On the other hand, the
assertion (iii)(c) does not extend to $3\times 4$ PPTES of rank
five, see Example \ref{Besk-Fam,KERNEL=finite}. Thus, there exist
PPTES in higher dimensions which cannot be constructed via the UPB
approach. So, the higher dimensional cases are  essentially
different from the two-qutrit case \cite{cd11JMP}. Finally the
assertion (iii)(d) does not extend to $3\times N$ PPTES of rank
$N+1$ when $N>3$. Indeed, such state may contain infinitely many
product vectors in the kernel, see Example \ref{Besk-Fam}.

Let $\s$ be an $M\times N$ PPT state of rank $N$. By Proposition
\ref{prop:PPTMxNrankN}, $\s$ is separable. Moreover, $\s$ is
SLOCC-equivalent to a state $\r$ given by Eq. (\ref{MxN-State})
where all $C_i$ are diagonal matrices. This fact follows from
\cite[Proposition 6 (ii)]{cd11JPA}, and will be used in several
proofs in this paper.

We need the following simple fact.
 \bl \label{le:Range}
Let $\r,\r'$ be bipartite states. If $\cR(\r')\se\cR(\r)$ then
$\cR(\r'_A)\se\cR(\r_A)$.
 \el
 \bpf
For small $\ve>0$ we have $\r-\ve\r'\ge0$. Hence,
$\r_A-\ve\r'_A\ge0$ and the assertion follows.
 \epf

As an application, we have the following fact.
 \bpp \label{prop:3x3rank4PPTES,samerange}
If the normalized states $\r$ and $\r'$ are $3\times3$ PPTES
of rank four with the same range, then $\r=\r'$.
 \epp

By Lemma \ref{le:Range}, we must have
$\cR(\r_A)=\cR(\r'_A)$ and $\cR(\r_B)=\cR(\r'_B)$.
Then the result follows from \cite[Theorem 22]{cd11JMP}.

We also need the concept of irreducibility for bipartite states
introduced in \cite[Definition 11]{cd11JPA}. We extend the
definition of A and B-direct sums to arbitrary linear operators.
 \bd
\label{def:reducibleirreducible} We say that a linear operator
$\r:\cH\to\cH$  is an {\em A-direct sum} of linear operators
$\r_1:\cH\to\cH$ and $\r_2:\cH\to\cH$, and we write
$\r=\r_1\oplus_A\r_2$, if
$\cR(\r_A)=\cR((\r_1)_A)\oplus\cR((\r_2)_A)$. (Note that we do
not require the ranges of $(\r_1)_A$ and $(\r_2)_A$ to be
orthogonal to each other.) A bipartite state $\r$
is {\em A-reducible} if it is an A-direct sum of two states;
otherwise $\r$ is {\em A-irreducible}. One defines similarly the
{\em B-direct sum} $\r=\r_1\oplus_B\r_2$, the {\em B-reducible} and
the {\em B-irreducible} states. We say that a state $\r$ is {\em
reducible} if it is either A or B-reducible. We say that $\r$ is
{\em irreducible} if it is not reducible. We write
$\r=\r_1\oplus\r_2$ if $\r=\r_1\oplus_A\r_2$ and
$\r=\r_1\oplus_B\r_2$, and in that case we say that $\r$ is a {\em
direct sum} of $\r_1$ and $\r_2$.
 \ed

The definitions of ``reducible'', ``irreducible'' and
``direct sum'' of two states are designed for use in the
bipartite setting and should not be confused with the usual
definitions of these terms where $\cH$ is not equipped
with the tensor product structure.
If $\r_1$ and $\r_2$ are states on the same Hilbert space, which
represents a bipartite quantum system, then it is straightforward
to check whether their sum is A-direct. However, if $\r_1$ and $\r_2$ act on two different Hilbert spaces representing two different bipartite
quantum systems, one may wish to embed these two
Hilbert spaces into a larger one, $\cH$, which also represents a
bipartite quantum system, such that the sum of $\r_1$ and $\r_2$
becomes an A-direct sum. This can be accomplished in many
different ways, but there is no natural or canonical way to
select such a construction. For that reason there is no operation
of ``forming'' the A-direct sum of $\r_1$ and $\r_2$, and in each case such a construction has to be explained in more details.
Of course, this warning applies also to B-direct sums.

Let $A=B+C$ where $B$ and $C$ are Hermitian matrices and $\rank
A=\rank B+\rank C$. Then it is easy to show that $A\ge0$ implies
that $B\ge0$ and $C\ge0$. Consequently, if $\r=\r_1\oplus_A\r_2$ or
$\r=\r_1\oplus_B\r_2$ with $\r_1$ and $\r_2$ Hermitian and $\r\ge0$,
then also $\r_1\ge0$ and $\r_2\ge0$.

Let us recall a related result \cite[Corollary 16]{cd11JPA} to
which we will refer in many proofs.
 \bl \label{le:reducible=SUMirreducible,SEP,PPT}
Let $\r=\sum_i\r_i$ be an A or B-direct sum of the states $\r_i$.
Then $\r$ is separable [PPT] if and only if each $\r_i$ is
separable [PPT]. Consequently, $\r$ is a PPTES if and only if each
$\r_i$ is PPT and at least one of them is entangled.
 \el

It follows from this lemma that any extreme state is irreducible. We
insert here a new lemma.
 \bl
 \label{le:RHOreducible=RHOPTreducible}
Let $\r_1$ and $\r_2$ be linear operators on $\cH$.

(i) If $\r=\r_1\oplus_B\r_2$, then $\r^\G=\r_1^\G\oplus_B\r_2^\G$.

(ii) If $\r_1$ and $\r_2$ are Hermitian and $\r=\r_1\oplus_A\r_2$,
then $\r^\G=\r_1^\G\oplus_A\r_2^\G$.

(iii) If a PPT state $\r$ is reducible, then so is $\r^\G$.
 \el
 \bpf
(i) follows from the fact that $(\s^\G)_B=\s_B$ for any state $\s$
on $\cH$, see Eq. \eqref{ea:rhoBGamma=rhoB}.

(ii) First observe that  $(\s^\G)_A=(\s_A)^T$ for any state $\s$
on $\cH$, see Eq. \eqref{ea:rhoAGamma=rhoGammaA}.
Then the assertion follows from the fact that $\cR(\s^T)=\cR(\s)^*$
for any Hermitian operator $\s$ on $\cH_A$.

(iii) follows immediately from (i) and (ii).
 \epf

Let $\r$ be any $M\times N$ state and $\ket{a}\in\cH_A$ a nonzero
vector. Then it is easy to verify that $\bra{a}\r\ket{a}\ne0$.
(Similarly, $\bra{b}\r\ket{b}\ne0$ for any
nonzero vector $\ket{b}\in\cH_B$.)
The following two assertions are equivalent to each other:

(i) $\rank\bra{a}\r\ket{a}=1$;

(ii) $\ket{a}\ox H\subseteq\ker\r$ for some hyperplane $H\subset\cH_B$.

Let us state the general extremality criterion which was
discovered recently by Leinaas, Myrheim and Ovrum \cite{lmo07} and
independently by Augusiak, Grabowski, Kus and Lewenstein
\cite{agk10}. We offer an enhanced version of this criterion and
give a short proof. The new assertion (ii) in this criterion plays
an essential role in the proof of Theorem \ref{thm:MxNrank(M+N-2)}.

 \bpp \label{prop:ExtCrit}
 {\em (Extremality Criterion)}
For a PPT state $\r$, the following assertions are equivalent to
each other.

(i) $\r$ is not extreme.

(ii)  There is a PPT state $\s$, not a scalar multiple of $\r$, such
that $\cR(\s)=\cR(\r)$ and $\cR(\s^\G)=\cR(\r^\G)$.

(iii) There is a Hermitian matrix $H$, not a scalar multiple of
$\r$, such that $\cR(H)\subseteq\cR(\r)$ and
$\cR(H^\G)\subseteq\cR(\r^\G)$.
 \epp
 \bpf (i) $\to$ (ii). We have $\r=\r_1+\r_2$ where $\r_1$ and $\r_2$
are non-parallel PPT states. We also have $\r^\G=\r_1^\G+\r_2^\G$.
Then the state $\s:=\r+\r_1$ is not a scalar multiple of $\r$ and
satisfies $\cR(\s)=\cR(\r)$ and $\cR(\s^\G)=\cR(\r^\G)$.

(ii) $\to$ (iii) is trivial.

(iii) $\to$ (i). It follows from (iii) that there exists $\ve>0$
such that $\r+tH\ge0$ and $\r^\G+tH^\G\ge0$ for $t\in[-\ve,\ve]$.
Then $\r_1=\r-\ve H$ and $\r_2=\r+\ve H$ are non-parallel PPT
states and $\r_1+\r_2=2\r$. Hence (i) holds.
 \epf

The equivalence of (i) and (ii) is a trivial consequence of the
description of the faces of the convex cone of non-normalized PPT states given in \cite{hk05}.

The following necessary condition for extremality was first
discovered by Leinaas, Myrheim and Ovrum \cite{lmo07}.
Our concise proof below is essentially the same as their proof.

 \bpp \label{prop:extremeNecCond}
Let $\r$ be a PPT state of birank $(r,s)$. If $r^2+s^2>M^2 N^2+1$
then $\r$ is not extreme.
 \epp
 \bpf Let $Q$ be the real vector spaces of all Hermitian
matrices of size $MN\times MN$. Denote by $Y$ $[Z]$ the subspace
of $Q$ consisting of all Hermitian matrices whose range is
contained in $\cR(\r)$ $[\cR(\r^\G)]$. Note that $\dim Y=r^2$,
$\dim Z=s^2$ and $\dim Q=M^2 N^2$. The subspace
$Z^\G:=\{H^\G:H\in Z\}\subseteq Q$ also has dimension $s^2$. We need to estimate the dimension of the subspace
$V:=\{H\in Y: H^\G\in Z\}$. Since $V=Y\cap Z^\G$, we have
 \bea
 \dim V &=& \dim (Y\cap Z^\G) \notag \\
    &\ge& \dim Y +\dim Z^\G - \dim Q \notag \\
    &=& r^2+s^2-M^2 N^2 > 1.
 \eea
Hence, the assertion (iii) of Proposition \ref{prop:ExtCrit}
holds, and so $\r$ is not extreme.
 \epf

For instance, when $M=2$ and $N=4$ we see immediately that there
are no $2\times4$ extreme states of birank $(6,6)$.

For a PPT state $\r$, if $\ket{a,b}\in\ker\r$ then
$\ket{a^*,b}\in\ker\r^\G$, see \cite[Lemma 5]{kck00}.
Thus the partial conjugation automorphism $\S\to\S$ maps $X_\r$
onto $X_{\r^\G}$, and so part (i) of the following lemma holds.

 \bl \label{le:rho,rhoGamma=delta}
Let $\r$ be a PPT state of rank $r$. Then

(i) $|X_{\r^\G}|=|X_\r|$ and $\dim X_{\r^\G}=\dim X_\r$;

(ii) if $\r$ is good and $r\le M+N-2$, then $\rank\r^\G\ge r$.
 \el
 \bpf We prove (ii). It follows from the hypothesis that
$\dim X_\r=M+N-2-r$. Assume that $\rank\r^\G<r$. Then
\begin{equation}
\dim X_{\r^\G}\ge M+N-2-\rank\r^\G>M+N-2-r=\dim X_\r,
\end{equation}
which contradicts (i).
 \epf


Finally, let us prove two basic facts about strongly extreme states.
 \bl \label{le:stronglyextreme}
Let $\s$ be a strongly extreme state.

(i) If $\r$ is a PPT state and $\cR(\r)\subseteq\cR(\s)$, then
$\r\propto\s$. In particular, a strongly extreme state is extreme.

(ii) If $\rank\s>1$ then $\cR(\s)$ is a CES.
 \el
 \bpf
(i) Since $\r+\s$ is PPT, we must have $\r+\s\propto\s$. Hence
$\r\propto\s$.

(ii) Assume that $\cR(\s)$ contains a product vector $\ket{a,b}$.
Then $\r=\proj{a,b}$ is a PPT state and $\cR(\r)\subseteq\cR(\s)$.
By (i) we have $\r\propto\s$, which contradicts the hypothesis that
$\rank\s>1$. Hence $\cR(\s)$ must be a CES.
 \epf

\section{Good and bad states}
\label{sec:goodsep}

We have divided the bipartite states into good and bad ones. Good states are of more interest since they share many good properties.
The main result of this section is the characterization of
good separable states, see Theorem \ref{thm:goodseparablestate}.
We give explicit expression for any good separable state by using the
product vectors contained in the range. Some preliminary results,
like Proposition \ref{prop:MxN,>delta=infinitelymany} and Lemma
\ref{le:GenPos,L<=M+N-2}, treat general vector subspaces of $\cH$
and will be useful later.
The proof of Proposition \ref{prop:MxN,>delta=infinitelymany}
is based on two facts from algebraic geometry for which we could
not find a reference.
Their proofs are given in the appendix.
We also show that all pure states are good.

\bd \label{def:VeryGood}
We say that a state $\r$ acting on $\cH=\cH_A\ox\cH_B$ is
{\em universally good} if it is good and remains good whenever
our $M\ox N$ system $\cH$ is embedded in a larger $M'\ox N'$ system.
 \ed

To take care of the trivial cases $M=1$ or $N=1$, we observe that
in these cases $\cP_{AB}=\S$ and so every subspace of $\cH$ is good,
and every state on $\cH$ is good.

Let us now show that all pure states are universally good.
 \bpp  \label{prop:puregood}
$(M,N\ge1)$ Every pure state is good and, consequently, it is
universally good.
 \epp
 \bpf
Let $\r=\proj{\Psi}$ be any pure state.
We may assume that $M,N>1$. Since $\rank\r=1$,
$K:=\ker\r$ is a hyperplane of $\cH$. After applying an ILO, we may
assume that $\ket{\Psi}=\sum_{i=0}^{r-1}\ket{ii}$, where
$r\le\min(M,N)$ is the Schmidt rank of $\ket{\Psi}$. For
$\ket{x}=\sum_i\x_i\ket{i}_A$ and $\ket{y}=\sum_j\eta_j\ket{j}_B$,
we have $\ket{x,y}\in K$ if and only if $\braket{\Psi}{x,y}=0$,
i.e., $\sum_{i=0}^{r-1}\x_i\eta_i=0$. Hence, $X_\r$ is an
irreducible hypersurface in $\S$ except when $r=1$ in which case it
has two irreducible components: $\x_0=0$ and $\eta_0=0$. Since
$\ket{00}\notin K$, we see that $\ket{0}\ox\cH_B\not\se K$.
Consequently, $\hat{K}$ and $\S$ intersect transversely at the point
represented by $\ket{01}$ and our claim is proved if $r>1$. If $r=1$
the above argument shows that the transversality condition is
satisfied by the component $\eta_0=0$. The other component can be
dealt with in the same manner.
 \epf

Thus the difference $M-\rank\r_A$ (and $N-\rank\r_B$) may be
arbitrarily large for good states $\r$.
On the other hand, we will now show that good PPTES must have
full local ranks (i.e., $\rank\r_A=M$ and $\rank\r_B=N$).
To do this we need a preliminary fact.
 \bl
 \label{le:good,rankrho>rankrhoA}
$( M,N\ge1)$ Let $\r$ be a good state of rank $r>\rank\r_A$.
Then $\rank\r_A=M$ and, consequently, $\r$ is not universally good:
it becomes bad when our $M\ox N$ system is embedded in any
larger system $M'\ox N'$ with $M'>M$.
 \el
 \bpf
The first assertion is trivial when $M=1$. Let $M>1$. Assume that
$\rank\r_A<M$. Since $H:=\cR(\r_A)^\bot\ox\cH_B\subseteq\ker\r$ is
nonzero, we have $X_\r\ne\es$. As $\r$ is good, it follows that
$r:=\rank\r\le M+N-2$ and $\dim X_\r=M+N-2-r$. Since
$X_\r\supseteq\hH\cap\S$ and $\dim\hH\cap\S=M+N-2-\rank\r_A$, we
have $\rank\r_A \ge r$, which contradicts the hypothesis. Thus
$\rank\r_A=M$ and the first assertion is proved.

The second assertion follows from the first.
 \epf

Thus, a product state $\r=\r_A \ox \r_B$ of rank bigger than one
is not universally good.

We now exhibit a link between goodness and distillability properties
of entangled bipartite states.
 \bt
 \label{thm:goodnondistillable}
If a good entangled state $\r$ is not distillable (e.g., if it
is PPT), then it must have full local ranks (i.e., $\r$ must be
an $M\times N$ state).
 \et
 \bpf Since $\r$ is not distillable, we have $\rank\r\ge\rank\r_A$
by Theorem \ref{thm:PPTMxNrank<M,N}.
If $\r$ is PPT, Proposition \ref{prop:PPTMxNrankN} implies
that $\rank\r\ne\rank\r_A$.
By \cite[Theorem 10]{cd11JPA}, this is also true when $\r$ is NPT.
Thus $\rank\r>\rank\r_A$ and
Lemma \ref{le:good,rankrho>rankrhoA} implies that $\r_A$
has full rank. Similarly, $\r_B$ has full rank.
 \epf

 \bpp
 \label{prop:MxN,>delta=infinitelymany}
Let $H$ be a vector subspace of $\cH$ of dimension $d$ containing
exactly $m$, $0\le m<\infty$, product vectors.

(i) Then $m\le\d$ and $d\le(M-1)(N-1)+1$.

(ii) If $m=\d$ then $d=(M-1)(N-1)+1$, $H$ is spanned by product
vectors and no proper subspace $V\ox W$ of $\cH$ contains $H$.

(iii) If $d\le(M-1)(N-1)$ then $m<\d$.
 \epp
 \bpf
It follows from Proposition \ref{prop:Intersection} that
$d\le(M-1)(N-1)+1$. By Proposition \ref{prop:David}, there exists a
vector subspace $H'\supseteq H$ of dimension $(M-1)(N-1)+1$ such
that $H'$ contains only finitely many, say $m'$, product vectors. By
the B\'{e}zout's theorem, we have $m'\le\d$. Since $m\le m'$, we also
have $m\le\d$. Thus (i) is proved.

If $m=\d$ then also $m'=\d$ and Theorem \ref{thm:PresekProjVar}
implies that $H'=H$, i.e., $d=(M-1)(N-1)+1$, and that $H$ is spanned
by product vectors. Thus (ii) is proved. The assertion (iii) follows
from (i) and (ii).
 \epf

This proposition will be used in the proofs of Theorems
\ref{thm:delta} and \ref{thm:LMS2010(ii),ker=finitePRO}, which
are our main results regarding Conjectures
\ref{conj:LMS2010highdimension} and \ref{conj:EkstrSt}.

Let us make a comment about the case $m=\d$. In that case $H$ has
a basis consisting of product vectors, say $\ket{a_i,b_i}$,
$i=1,\ldots,d$. If $V$ $[W]$ is the subspace of $\cH_A$ $[\cH_B]$
spanned by the $\ket{a_i}$ $[\ket{b_i}]$, then each
$\ket{a_i,b_i}\in V\ox W$.
It follows that $V=\cH_A$ and $W=\cH_B$.
Thus, we have the following corollary.

 \bcr \label{cor:KerGoodStates}
Let $\r$ be a good $M\times N$ state of rank $M+N-2$. Then
$\ker\rho$ is spanned by product vectors and $|X_\r|=\d$.
If $\{\ket{a_i,b_i}\}$ is any basis of $\ker\r$
consisting of product vectors, then the $\ket{a_i}$ span $\cH_A$ and
the $\ket{b_i}$ span $\cH_B$.
 \ecr

The hypothesis that $\r$ is good is essential as the following
example shows.
\begin{example} \label{eg:3x3rank4,kerNEprodspan} $(M=N=3)$
{\rm Consider the $3\times3$ separable state
$\r=\sum^2_{i=0}\proj{ii}+\proj{a,b}$ of rank four, where
$\ket{a}=\ket{1}_A+\ket{2}_A$ and $\ket{b}=\ket{1}_B+\ket{2}_B$. One
can check that a product vector $\ket{x,y}$ belongs to $\ker\r$ if
and only if the vectors $\ket{x}=\sum_i \x_i\ket{i}$ and
$\ket{y}=\sum_i \eta_i\ket{i}$ satisfy the equations
 \bea
\x_0\eta_0=\x_1\eta_1=\x_2\eta_2=\x_1\eta_2+\x_2\eta_1=0.
 \eea
Hence such $\ket{x,y}$ belongs to one of the subspaces
$\ket{0}\ox\ket{0}^\perp$ or $\ket{0}^\perp\ox\ket{0}$. Since these
subspaces are contained in $\ker\r$, $\r$ is bad. As $\ker\r$ has
dimension five, it is not spanned by product vectors.
We mention that $\r^\G=\r$ in this example. $\square$ }
\end{example}

The projective variety $X_{\r}$ of the separable state $\r$ in
this example has only two irreducible components, the Segre
varieties of the subspaces $\ket{0}\ox\ket{0}^\perp$ and
$\ket{0}^\perp\ox\ket{0}$.

We can extend this observation to any separable state $\r$ of rank
$r$. We can write $\r$ as a sum of pure product states
 \bea \label{SepState}
\r=\sum_{i=1}^m \proj{a_i,b_i}, \quad m\ge r.
 \eea
For any subsets $P,Q\se I:=\{1,\ldots,m\}$ we set
 \bea \label{VJiWJ}
V_P=\{\ket{a_j}_{j\in P}\}^\perp\subseteq\cH_A, \quad
W_Q=\{\ket{b_k}_{k\in Q}\}^\perp\subseteq\cH_B.
 \eea

For simplicity, let us denote by $\S_{P,Q}$ the Segre variety of the
tensor product $V_P\ox W_Q$. (If $V_P=0$ or $W_Q=0$ then
$\S_{P,Q}=\emptyset$.) It is obvious that if $P\se P'\se I$ and
$Q\se Q'\se I$, then $\S_{P',Q'}\se\S_{P,Q}$.

 \bt
 \label{thm:irreducible=segre}
Let $\r$ be a separable state given by Eq. (\ref{SepState}). Then
any irreducible component of $X_{\r}$ is one of the Segre varieties
$\S_{P,Q}$, where $(P,Q)$ runs through all partitions of the index
set $I=\{1,\ldots,m\}$.
 \et
 \bpf
Our first claim is that if $I=P\cup Q$, then $V_P\ox W_Q\se\ker\r$
and so $\S_{P,Q}\se X_\r$. For any $i\in I$ we have $i\in P$ or
$i\in Q$, say $i\in P$. By definition of $V_P$, $\ket{a_i}$ is
orthogonal to $V_P$, and so $\ket{a_i,b_i}$ is orthogonal to $V_P\ox
W_Q$. As the $\ket{a_i,b_i}$ span $\cR(\r)$, our first claim
follows.

Our second claim is that for any product vector $\ket{a,b}\in\ker\r$
there exists a partition $(P,Q)$ of $I$ such that $\ket{a,b}\in
V_P\ox W_Q$. To prove this claim, let $P$ $[Q]$ to be the set of
indexes $j$ $[k]$ such that $\braket{a}{a_j}=0$
$[\braket{b}{b_k}=0]$. Since
$\braket{a}{a_i}\braket{b}{b_i}=\braket{a,b}{a_i,b_i}=0$ for each
$i\in I$, we have $P\cup Q=I$. By replacing $Q$ with $Q\setminus P$,
we obtain a partition of $I$ and our second claim follows.

Hence, the variety $X_\r$ is the union of the Segre subvarieties
$\S_{P,Q}$ where $(P,Q)$ runs through all partitions of $I$. The
assertion of the theorem follows because there are only finitely
many partitions $(P,Q)$ of $I$ and each Segre variety $\S_{P,Q}$ is
irreducible.
 \epf

We need the following lemma, where we use the concept of ``general
position'' (see Definition \ref{def:GenPos}).

 \bl \label{le:GenPos,L<=M+N-2}
Let $V\subseteq\cH$ be a subspace spanned by the product vectors
$\ket{a_i,b_i}$, $i=1,2,\ldots,L$,  in general position. If $L\le
M+N-2$ then the $\ket{a_i,b_i}$ are linearly independent and any
product vector in $V$ is a scalar multiple of some $\ket{a_i,b_i}$.
 \el
 \bpf
We may assume that $M\le N$. The proof is by induction on $L$. Both
assertions are true if $L=1$. Now let $L>1$. By the induction
hypothesis, the vectors $\ket{a_i,b_i}$, $1\le i<L$, are linearly
independent and $\ket{a_L,b_L}$ is not their linear combination.
Thus the vectors $\ket{a_i,b_i}$, $1\le i\le L$, are linearly
independent. It remains to prove the second assertion.

Suppose there exists a product vector $\ket{a,b}\in V$ which is not
a scalar multiple of any $\ket{a_i,b_i}$. We have $\ket{a,b}=\sum_i
\x_i \ket{a_i,b_i}$, $\x_i\in\bC$. The induction hypothesis implies
that all $\x_i\ne0$. Assume that $L=N$. Since the $\ket{a_i,b_i}$
are in general position, the vectors $\ket{b_1},\ldots,\ket{b_N}$
are linearly independent. As $\ket{a,b}$ is a product vector, it
follows that each of the vectors $\ket{a_1},\ldots,\ket{a_N}$ must
be a scalar multiple of $\ket{a}$. Thus we have a contradiction, and
we conclude that $L>N$.

Since $\{\ket{b_1},\ldots,\ket{b_N}\}$ is a basis of $\cH_B$, we
have
\begin{eqnarray}
\ket{b_i} &=& \sum_{j=1}^N \eta_{ij} \ket{b_j},\quad
\eta_{ij}\in\bC, \quad N<i\le L; \\  \label{eq:LinZav} \ket{a,b} &=&
\sum^{N}_{j=1}\bigg(\x_j\ket{a_j}+
\sum^L_{i=N+1}\x_i\eta_{ij}\ket{a_i}\bigg)\ox\ket{b_j}.
\end{eqnarray}
As $\x_1,\ldots,\x_{N}$ are nonzero, Eq. (\ref{eq:LinZav}) implies
that the vectors $\ket{a_1},\ldots,\ket{a_N}$ belong to the subspace
spanned by the $\ket{a_i}$ with $N<i\le L$ and $\ket{a}$. Since the
dimension of this subspace is at most $L-N+1\le M-1$ and $M\le N$,
we conclude that $\ket{a_1},\ldots,\ket{a_M}$ are linearly
dependent. This contradicts our hypothesis, and proves that the
second assertion is also valid.
 \epf

We can now characterize the good separable states.
 \bt  \label{thm:goodseparablestate}
$(M,N\ge1)$ Let $\r$ be a separable state of rank $r$.

(i) If $r\le M+N-2$ then $\r$ is good if and only if
$\r=\sum^r_{i=1}\proj{a_i,b_i}$, where the product vectors
$\ket{a_i,b_i}$, $i=1,\ldots,r$, are in general position.

(ii) If $r>M+N-2$ then $\r$ is good if and only if
$\r=\sum^m_{i=1}\proj{a_i,b_i}$ and, for any partition $I=P\cup Q$
of the index set $I=\{1,\ldots,m\}$, either the $\ket{a_j}$,
$j\in P$, span $\cH_A$ or the $\ket{b_k}$, $k\in Q$, span
$\cH_B$.

(iii) If $\r$ is good then so is $\r^\G$.
 \et
 \bpf
(i) \textit{Necessity}. Let $\r$ be given by Eq. (\ref{SepState})
where the $\ket{a_i,b_i}$ are pairwise non-parallel. We may assume
that the $\ket{a_i,b_i}$, $i=1,\ldots,r$ span $\cR(\r)$. Assume that
these $r$ product vectors are not in general position, say
$\ket{a_1},\ldots,\ket{a_n}$ are linearly dependent. Set
$P=\{1,\ldots,n\}$ and $Q=\{n+1,\ldots,r\}$. Define the subspaces
$V_P$ and $W_Q$ as in Eq. (\ref{VJiWJ}). We have $\dim V_P\ge
M-n+1$, $\dim W_Q\ge n+N-r$ and $V_P\ox W_Q\se\ker\r$. Hence $\dim
X_\r\ge\dim\S_{P,Q}>M+N-r-2$, which contradicts the hypothesis that
$\r$ is good. Thus, the product vectors $\ket{a_i,b_i}$, $i\le r$,
must be in general position. Now Lemma \ref{le:GenPos,L<=M+N-2}
implies that $m=r$.

\textit{Sufficiency}. We may assume that $M,N>1$. By Theorem
\ref{thm:irreducible=segre}, every irreducible component of $X_\r$
is the Segre variety $\S_{P,Q}$ for some partition $(P,Q)$ of
$\{1,\ldots,r\}$. We may assume that $|P|<M$ and $|Q|<N$ since
otherwise $\S_{P,Q}=\es$. Note that then the $\S_{P,Q}$ have
dimension $M+N-2-r$, and so $\dim X_\r=M+N-2-r$. It remains to
verify the transversality condition. We choose $\ket{a}\in V_P$ and
$\ket{b}\in W_Q$ such that $\braket{a_k}{a}\ne0$ for $k\in Q$ and
$\braket{b_j}{b}\ne0$ for $j\in P$. We have to show that
$\ker\r+S_{a,b}=\cH$. For this it suffices to show that $\ker\r\cap
S_{a,b}\se V_P\ox\ket{b}+\ket{a}\ox W_Q$. Let $\ket{\psi}=
\ket{a,y}+\ket{x,b}\in\ker\r$. Then $\r\ket{\psi}=0$, which gives
the equations
$\braket{b_i}{b}\braket{a_i}{x}+\braket{a_i}{a}\braket{b_i}{y}=0$
for $i=1,\ldots,r$. Since $\braket{b_i}{b}=0$ for $i\in Q$ and
$\braket{a_i}{a}=0$ for $i\in P$, we get the equations
$\braket{a_j}{x}=0$ for $j\in P$ and $\braket{b_k}{y}=0$ for $k\in
Q$. Thus $\ket{\psi}\in V_P\ox\ket{b}+\ket{a}\ox W_Q$. Hence, the
transversality condition is satisfied and so $\r$ is good.

(ii) This follows immediately from Theorem
\ref{thm:irreducible=segre}.

(iii) When $r\le M+N-2$, it follows from (i) that
$\r^\G=\sum^r_{i=1}\proj{a_i^*,b_i}$, where the $\ket{a_i^*,b_i}$
are in general position. By Lemma \ref{le:GenPos,L<=M+N-2} we have
$\rank\r^\G=r$, and (i) shows that $\r^\G$ is good.

If $r>M+N-2$ then $X_\r=\es$ and, by Lemma \ref{le:rho,rhoGamma=delta},
also $X_{\r^\G}=\es$. Hence, $\r^\G$ is good.
 \epf

Although, for separable $\r$, either both $\r$ and $\r^\G$ are good
or both bad, they may have different ranks in case (ii). A
well-known example is the separable two-qubit Werner state $\r=I\ox
I + \sum^1_{i,j=0}\ketbra{ij}{ji}$. It is good since $X_\r=\es$, but
its birank is $(3,4)$.

As a simple corollary, we show that good separable states in (i)
indeed satisfy the degree formula \eqref{ea:DegreeFormula}. There
are ${r \choose k}$ partitions $(P,Q)$ of $\{1,\ldots,r\}$ such that
$|P|=k$. For such partitions $(P,Q)$, the degree of $\S_{P,Q}$ is
${M+N-2-r \choose M-1-k}$. Hence, the sum of the degrees of all
irreducible components of $X_\r$ is the left hand side of the
identity
 \bea
\sum^{M-1}_{k=r-N+1}{r \choose k}\cdot{M+N-2-r \choose
M-1-k}
 ={M+N-2 \choose M-1}.
 \eea
It is easy to verify this identity, and so
Eq. \eqref{ea:DegreeFormula} is satisfied.

We can now characterize the universally good PPT states.
 \bt
 \label{thm:good,PPT}
A PPT state $\r$ is universally good if and only if
$\r=\sum_i\proj{a_i,b_i}$ where
the $\ket{a_i}$ and the $\ket{b_i}$ are linearly independent.
 \et
 \bpf
Let $r_a=\rank\r_A$, $r_b=\rank\r_B$ and $r=\rank\r$.

\textit{Necessity}. Suppose $\r$ is universally good.
By Lemma \ref{le:good,rankrho>rankrhoA} we have $r\le\min(r_a,r_b)$.
Since $\r$ is PPT, Theorem \ref{thm:PPTMxNrank<M,N} shows that
$r\ge\max(r_a,r_b)$.
Hence, we have $r_a=r_b=r$ and the assertion follows from
Proposition \ref{prop:PPTMxNrankN}.

\textit{Sufficiency}. When $r=1$, the claim
follows from Proposition \ref{prop:puregood}. When $r>1$,
Theorem \ref{thm:goodseparablestate} (ii) applies.
 \epf

So far, pure entangled states are the only known NPT states, with
$\rank\r\le\min(\rank\r_A,\rank\r_B)$, which are universally good.
Constructing more examples of such states is an interesting problem.

\section{$M\times N$ PPT states of rank $M+N-2$}
\label{sec:generic}

This section is split into two subsections. In the first subsection
we prove the basic property of $M\times N$ PPT states $\r$ of rank
$M+N-2$, namely that if $X_\r$ is a finite set then $|X_\r|=\d$.
See Theorem \ref{thm:delta} below for a stronger version of this
result.
In the second subsection we prove that part (iii) of
Conjecture \ref{conj:LMS2010highdimension} and
Conjecture \ref{conj:EkstrSt} are valid in the good case.

\subsection{Product vectors in the kernel}

Motivated by Conjecture \ref{conj:LMS2010highdimension}, we
shall prove a general theorem about arbitrary $M\times N$ PPT
states. The proof is an extension of the proof of
\cite[Theorem 20]{cd11JMP}. We recall that the Segre variety
$\S=\S_{M-1,N-1}$ and the number $\d$ were defined in Section
\ref{sec:introduction}, see formula (\ref{jed:delta}).
Note that if the kernel of a state $\r$ contains a 2-dimensional
subspace $V\ox W$, then the variety $X_\r$ contains a projective
line and so $\dim X_\r\ge1$.

 \bt \label{thm:delta}
If $\r$ is an $M\times N$ PPT state of rank $r$ such that
$\ker\r$ contains no 2-dimensional subspace $V\ox W$, then either $r=M+N-2$ and $|X_\r|=\d$ or $r>M+N-2$ and $|X_\r|<\d$.
 \et
 \bpf If $K=\ker\r$ is a CES, then $r>M+N-2$ and the assertion
of the theorem holds. Thus we may assume that $K$ contains a
product vector. We choose an arbitrary product vector in $K$.
By changing the o.n. bases of $\cH_A$ and $\cH_B$,
we may assume that the chosen product vector is $\ket{00}$.
By using Eq. (\ref{MxN-State}), we may assume that
$\r=C^\dag C$, where $C=[C_0~C_1~\cdots~C_{M-1}]$ and the $C_i$
are $r\times N$ matrices.
Since $\ket{00}\in K$, the first column of $C_0$ is 0. The
hypothesis (with $\dim V=1$ and $\dim W=2$) implies that
$\rank\bra{a}\r\ket{a}\ge N-1$ for all nonzero vectors
$\ket{a}\in\cH_A$.
As $\bra{0}_A\r\ket{0}_A=C_0^\dag C_0$, the block $C_0$ must
have rank $N-1$, and so we may assume that
 \bea
C_0= \left[\begin{array}{cc}
0 & I_{N-1} \\
0 & 0
\end{array}\right]; \quad
C_i= \left[\begin{array}{cc}
u_i & * \\
v_i & *
\end{array}\right],~ 0<i<M,
 \eea
where $u_i\in\bC^{N-1}$ and $v_i\in\bC^{r-N+1}$ are column vectors.

Observe that the first entry of the matrix $\r$ is 0.
Since $\r^\G\ge0$, the first row of $\r^\G$ must be 0.
We deduce that $u_i=0$ for $i>0$.
The hypothesis (this time with $\dim V=2$ and $\dim W=1$)
implies that the first columns of the $C_i$, $0<i<M$, must be
linearly independent. In particular, we must have
$r-(N-1)\ge M-1$, i.e., $r\ge M+N-2$.

Let $\{e_i:1\le i<M\}$ be the standard basis of $\bC^{r-N+1}$.
By using an ILO on system A, we may assume that $v_i=e_i$ for
$0<i<M$. Thus we have
 \bea
C_i= \left[\begin{array}{cc}
0 & * \\
e_i & *
\end{array}\right],~ 0<i<M.
 \eea

The range of $\r$ is the subspace of dimension $r$ spanned by the
vectors $\ket{\psi_i}$, $i=1,\ldots,r$, given by the columns of
$C^\dag$. Each of these columns can be split into $N$ pieces of
height $M$ and the pieces arranged in natural order to form an
$M\times N$ matrix. By using this matrix notation, we have
 \bea
\ket{\psi_j} &=& \left[\begin{array}{cc}
0 & f_j^T \\
0 & B_j
\end{array}\right],~ j=1,\ldots,N-1; \\
\ket{\psi_{N+i-1}} &=& \left[\begin{array}{cc}
0 & 0 \\
e_i & B_{N+i-1}
\end{array}\right],~ i=1,\ldots,M-1; \\
\ket{\psi_{N+i-1}} &=& \left[\begin{array}{cc}
0 & 0 \\
0 & B_{N+i-1}
\end{array}\right],~ i=M,\ldots,r-N+1,
 \eea
where $\{f_j\}$ is the standard basis of $\bC^{N-1}$ and the
$B_k=[b^k_{ij}]$ are $(r-N+1)\times(N-1)$ matrices.

Let $\hat{K}$ be the projective space associated to $K$. We
introduce the homogeneous coordinates $\x_{ij}$ for the projective
space $\cP_{AB}$ associated to $\cH$: If
$\ket{\psi}=\sum_{i=0}^{M-1}\sum_{j=0}^{N-1}\a_{ij}\ket{ij}$ then
the homogeneous coordinates of the corresponding point
$\ket{\psi}\in\cP_{AB}$ are $\x_{ij}=\a_{ij}$.

We claim that $\dim X_\r=0$, i.e., $X_\r$ is a finite set.
To prove this claim,
we shall use the affine chart defined by $\x_{00}\ne0$ which contains
the chosen point $P=\ket{00}$.
We introduce the affine coordinates $x_{ij}$, $(i,j)\ne(0,0)$,
in this affine chart by setting $x_{ij}=\x_{ij}/\x_{00}$.
Thus $P$ is the origin, i.e., all of its affine coordinates $x_{ij}=0$.
Since $\ker\r=\cR(\r)^\perp$, the subspace $\hat{K}$ is
the zero set of the ideal $J_1$ generated by the $r$ linear polynomials
on the left hand side of the equations:
 \bea
\label{lin-pol-a} && x_{0k}+\sum_{i=1}^{M-1} \sum_{j=1}^{N-1}
(b^k_{ij})^* x_{ij}=0,
\quad k=1,\ldots,N-1; \\
\label{lin-pol-b} &&
x_{k0}+\sum_{i=1}^{M-1}\sum_{j=1}^{N-1}(b^{N+k-1}_{ij})^* x_{ij}=0,
\quad k=1,\ldots,M-1; \\
\label{lin-pol-c} && \sum_{i=1}^{M-1} \sum_{j=1}^{N-1}
(b^{N+k-1}_{ij})^* x_{ij}=0, \quad k=M,\ldots,r-N+1. \eea The piece
of $\S$ contained in our affine chart consists of all $M\times N$
matrices
 \bea
\left[\begin{array}{cccc}
1 & x_{01} & x_{02} & \cdots \\
x_{10} & x_{11} & x_{12} & \\
x_{20} & x_{21} & x_{22} & \\
\vdots & & &
\end{array}\right]
 \eea
of rank one. It is the zero set of the ideal $J_2$ generated by the
$(M-1)(N-1)$ quadratic polynomials $x_{ij}-x_{i0}x_{0j}$, $1\le
i<M$, $1\le j<N$. By substituting $x_{ij}=x_{i0}x_{0j}$ $(i,j>0)$
into Eqs. (\ref{lin-pol-a}-\ref{lin-pol-b}), we obtain a system of
$M+N-2$ equations in $M+N-2$ variables to which we can apply Theorem
(1.16) of Mumford \cite{mfd76}. By that theorem, the singleton set
$\{P\}$ is an irreducible component of the affine variety defined by
the $M+N-2$ equations mentioned above. This remains true if we
enlarge this set of equations with those in (\ref{lin-pol-c})
because all of them vanish at the origin. We conclude that $\{P\}$
is also an irreducible component of $X_\r$. Since the point $P$ was
chosen arbitrarily in $X_\r$, our claim is proved.

If $r>M+N-2$ then the fact that $X_\r$ is a finite set and
Proposition \ref{prop:MxN,>delta=infinitelymany} (iii) imply that
$\ker\r$ contains at most $\d-1$ product vectors. It remains to
consider the case $r=M+N-2$. Note that now the set of
equations (\ref{lin-pol-c}) is empty.

Next we claim that the intersection multiplicity of $\hat{K}$ and $\S$
at the point $P=\ket{00}$ is 1. The computation of this multiplicity
is carried out in the local ring, say $R$, at the point $P$. This
local ring consists of all rational functions $f/g$ such that $g$
does not vanish at the origin, i.e., $f$ and $g$ are polynomials
(with complex coefficients) in the affine coordinates $x_{ij}$ and
$g$ has nonzero constant term. By expanding these rational functions
in the Taylor series at the origin, one can view $R$ as a subring of
the power series ring $\bC[[x_{ij}]]$ in the $MN-1$ affine
coordinates $x_{ij}$. We denote by $\gm$ the maximal ideal of $R$
generated by the $x_{ij}$.

The quotient space $\gm/\gm^2$ is a vector space of dimension $MN-1$
with the images of the $x_{ij}$ as its basis. It is now easy to see
that the images of the generators of $J_1$ and $J_2$ also span the
space $\gm/\gm^2$. Hence, by Nakayama's Lemma (see \cite[p.
225]{Cox}) we have $J_1+J_2=\gm$. Consequently,
$R/(J_1+J_2)\cong\bC$ and so our claim is proved.

Recall that we chose in the beginning an arbitrary product vector in
$\ker\r$ and by changing the coordinates we were able to assume that
this product vector is $\ket{00}$. Since the intersection
multiplicity is invariant under coordinate changes, this means that
we have shown that the intersection multiplicity is 1 at each point
of $X_\r$. By the B\'{e}zout's theorem the sum of the multiplicities at
all intersection points is $\d$, and since all of the multiplicities
are equal to 1 we conclude that $|X_\r|=\d$.
This concludes the proof.
 \epf

By Theorem \ref{thm:delta}, an $M\times N$ PPT state $\r$ of rank
$M+N-2$ is good if and only if $|X_\r|<\infty$.
The analogous assertion for NPT states is false.
For counter-examples see the proof of \cite[Theorem 10]{cd11JMP}.

By Proposition \ref{prop:extremeNecCond} we have
$\cE_{MN}^{M,N}=\es$.

 \bcr
 \label{cor:upperbound}
We have $\cE_{MN-1}^{M,N}=\es$.
 \ecr
 \bpf
Assume that there exists $\r\in\cE_{MN-1}^{M,N}$ and let
$s=\rank\r^\G$.
Obviously, we must have $MN>6$. Since $\r^\G$ is extreme and
$\cE_2=\cE_3=\es$, we must have $s>3$.
By Proposition \ref{prop:extremeNecCond} we have
$s^2\le M^2N^2+1-(MN-1)^2=2MN$.
If $M=N=3$, the only possibility is $s=4$.
Then Theorem \ref{thm:3x3PPTstates} gives a contradiction.
Thus, the case $M=N=3$ is ruled out.
In the remaining cases we have $2MN\le(M+N-2)^2$ and so $s\le M+N-2$.
It follows from Theorem \ref{thm:delta} that $|X_{\r^\G}|\ge\d$.
By Lemma \ref{le:rho,rhoGamma=delta}, we have $|X_\r|\ge\d$.
Since $\d>1$ and $\dim \ker\r=1$, we have a contradiction.
 \epf

Thus if $\cE_r^{M,N}\ne\es$ then $r\le MN-2$. This upper bound
is sharp in the sense that $\cE_{MN-2}^{M,N}$ may be nonempty.
Indeed there is a family of  $3\times3$ edge states $\r$  of birank
$(5,7)$ depending on a real parameter $\theta$ \cite{ko12}.
By using the extremality criterion, one can show that the state $\r$
is extreme when $\cos\theta=9/14$.

Assuming that part (ii) of Conjecture \ref{conj:LMS2010highdimension}
is valid, Proposition \ref{prop:extremeNecCond} gives a stronger
(at least for large $M,N$) upper bound $r^2\le M^2N^2+1-(M+N-2)^2$.
Without this conjecture, this stronger bound is valid for $M\le4$.

\subsection{Good states }

Next we prove that part (iii) of
Conjecture \ref{conj:LMS2010highdimension} and
Conjecture \ref{conj:EkstrSt} hold for good states.

  \bt
 \label{thm:LMS2010(ii),ker=finitePRO}
$(M,N>2)$ Let $\r$ be a good $M\times N$ PPT state of rank $M+N-2$.

(i) Then $\r$ is irreducible.

(ii) $\r^\G$ is also a good $M\times N$ PPT state of rank $M+N-2$.

(iii) If $\r$ is extreme, then $\r$ and $\r^\G$ are strongly extreme.

(iv) If $\r_1$ is an $M_1\times N_1$ state (acting on $\cH$) of
rank $r_1>N_1$ and $\cR(\r_1)\se\cR(\r)$, then $N_1=N$.

(v) If $\r$ is entangled and $\cE^{M,N}_r=\emptyset$ for
$1<r<M+N-2$, then $\r$ is extreme.
 \et
 \bpf
(i) Suppose $\r$ is reducible, say $\r=\r_1\op_B\r_2$. Let
$V_i=\cR((\r_i)_A)$ and $W_i=\cR((\r_i)_B)$, $i=1,2$. By Lemma
\ref{le:reducible=SUMirreducible,SEP,PPT}, $\r_i$ is an $M_i\times
N_i$ PPT state of rank $r_i$ where $M_i=\dim V_i$ and $N_i=\dim
W_i$. By using \cite[Proposition 15]{cd11JPA}, we may assume that
$W_1\perp W_2$. Let the state $\r'_i$ be the restriction of $\r_i$
to the subspace $V_i\ox W_i$. Since $V_1\ox W_1$ is orthogonal to
$V_2\ox W_2$, we have $\ker\r'_i\subseteq\ker\r$, and so $\ker\r'_i$
contains only finitely many product vectors. Then Proposition
\ref{prop:Intersection} implies that $M_i+N_i-2\le r_i$. Since
$N_1+N_2=N$ and $r_1+r_2=M+N-2$, it follows that $M_1+M_2\le
M+2<2M$. Thus we may assume that say $M_1\le M-1$. Since
$V_1^\perp\ox W_1\subseteq\ker\r$, we must have $\dim V_1^\perp\ox
W_1=1$, i.e., $M_1=M-1$ and $N_1=1$. Since $r_1\le M_1N_1=M-1$,
Theorem \ref{thm:PPTMxNrank<M,N} implies that $r_1=M-1$ and so
$r_2=N-1=N_2$. By Proposition \ref{prop:PPTMxNrankN}, $\r_2$ is a
sum of $N-1$ pure product states $\ket{a_i,b_i}$, $i=1,\ldots,N-1$.
Since $N_1=1$, there exists a nonzero vector $\ket{b}\in W_1^\perp$
which is orthogonal to all $\ket{b_i}$ with $i<N-1$. Then
$\ket{a_{N-1}}^\perp\ox\ket{b}\subseteq\ker\r$ which contradicts the
hypothesis of the theorem. Hence, the assertion (i) is proved.

(ii) By Eq. (\ref{prop:MxN,>delta=infinitelymany}), $\r^\G$ is
an $M\times N$ state.
By Theorem \ref{thm:delta}, $\ker\r$ contains exactly $\d$
product vectors. By Lemma \ref{le:rho,rhoGamma=delta},
$\ker(\r^\G)$ also  contains exactly $\d$ product vectors.
By Proposition \ref{prop:MxN,>delta=infinitelymany}, $\r^\G$
has rank $M+N-2$.

(iii) In view of (ii), it suffices to prove this assertion for $\r$
only. Let $\s$ be a PPT state such that $\cR(\s)=\cR(\r)$. By (ii),
all three states $\r^\G$, $\s^\G$ and $\r^\G+\s^\G$ have rank
$M+N-2$, and so they must have the same range. By Proposition
\ref{prop:ExtCrit}, $\s\propto\r$. Thus $\r$ is strongly extreme.

(iv) The assertion is true if $\r_1\propto\r$. Hence, we shall assume
that this is not the case. Without any loss of generality, we may
assume that $\r=\r_1+\r_2$ where $\r_2$ is an $M_2\times N_2$ state
of rank $r_2$.

Assume that $N_1<N$. By using Eq. (\ref{MxN-State}), we can write
the matrix of $\r$ as $\r=C^\dag C$, where
$C=[C_0~C_1~\cdots~C_{M-1}]$ and the matrices $C_i$, of size
$(r_1+r_2)\times N$, are block-triangular
 \begin{equation} \label{Blok-Mat-Lin}
C_i=\left[\begin{array}{cc}
    C_{i0} & 0
 \\ C_{i1} & C_{i3} \end{array}\right]
 \end{equation}
with $C_{i0}$ of size $r_1\times N_1$. The top $r_1$ [bottom $r_2$]
rows of $C$ represent $\r_1$ $[\r_2]$. Consequently, the matrix
$[C_{00}~C_{10}~\cdots~C_{M-1,0}]$ has rank $r_1$. Let $R$ be the
rank of the matrix $[C_{03}~C_{13}~\cdots~C_{M-1,3}]$.  Since $C$
has rank $M+N-2$, it is clear that $M+N-2\ge r_1+R$. We can choose a
unitary matrix $U$ such that the top $r_2-R$ rows of
$U[C_{03}~C_{13}~\cdots~C_{M-1,3}]$ are 0. Thus we may assume that
 \bea
C_i =
 \left[\begin{array}{cc}
    C_{i0}  & 0
 \\ C_{i11} & 0
 \\ C_{i12} & C_{i3}'
 \end{array}\right],
 \eea
where $C_{i3}'$ is of size $R\times (N-N_1)$.

Consider the PPT state $\s$ of rank $R$ defined by $\s=(C')^\dag
C'$, where $C'=[C_{03}'~C_{13}'~\cdots~C_{M-1,3}']$. Since $\s$ is
PPT and $\ker\r$ contains only finitely many product vectors, we
have $R\ge\rank\s_A \ge M-1$. Moreover, if $\rank\s_A=M-1$ then we
must have $N_1=N-1$. In that case we have
 \bea
 N_1=N-1\ge M+N-2-R\ge r_1 > N_1.
 \eea
which is a contradiction. We conclude that $\rank\s_A = M$.

Since $\s_B$ is a principal submatrix of $\r_B$, we have
$\rank\s_B=N-N_1$ and so
 \bea
\rank\s_A+\rank\s_B-2= M+(N-N_1)-2 \ge r_1+R-N_1 > R.
 \eea
Hence by Theorem \ref{thm:delta},
$\left(\cR(\s_A)\ox\cR(\s_B)\right)\cap\ker\s$ contains infinitely
many product vectors. As this subspace is contained in $\ker\r$, we
have again a  contradiction. Thus we have proved that $N_1=N$.

(v) Suppose $\r$ is not extreme. Then $\r=\sum_{i=1}^k\r_i$ with
pairwise non-parallel $\r_i\in\cE$ and, say, $\r_1$ entangled.
By Proposition \ref{prop:PPTMxNrankN}, $r_1:=\rank\r_1$ is bigger
than any of the two local ranks of $\r_1$.
Hence, by (iv), $\r_1$ must be an $M\times N$ state.
The hypothesis of (v) implies that $r_1=M+N-2$ and (iv) shows
that $\r_1$ is strongly extreme.
As $\cR(\r_2)\se\cR(\r)=\cR(\r_1)$, it follows that $\r_2\propto\r_1$
which is a contradiction.
 \epf

The assertion (ii) of Theorem \ref{thm:LMS2010(ii),ker=finitePRO}
may fail if $\r$ is bad. In the following example $\r$ is a
reducible PPTES. For another example with $\r$ reducible and
separable see Example \ref{eg:3x3birank(4,5)}.

\begin{example} $(M=N=4)$ \label{ex:M=N=4}
{\rm Consider the $4\times4$ reducible state $\r=\proj{00}\oplus\s$
of rank six, where $\s$ is a $3\times3$ edge state of rank five.
Such $\s$ may have the birank $(5,l)$ where $5\le l\le8$
\cite{hk05,kkl11}. Thus the birank of $\r$ is $(6,l+1)$, and so
$\rank\r<\rank\r^\G$ if $l\ne5$. As $\ket{0,i}\in\ker\r$ for
$i=1,2,3$, $\r$ is bad. \hfill $\square$ }
\end{example}

The assertion (iii) of
Theorem \ref{thm:LMS2010(ii),ker=finitePRO} does not hold when
$\rank\r>M+N-2$. For example the kernel of the $3\times3$ edge state
of rank five constructed in \cite[Sec. II]{clarisse06} has dimension
four but it contains only two product vectors. On the other hand, its
range contains a product vector so it is not strongly extreme.
(It is known and easy to check that this state is extreme.)

We shall now strengthen the assertion of Theorem \ref{thm:delta}
in the case $r<M+N-2$.

 \bt \label{thm:(M+N-r-1)x1=1x(M+N-r-1)ppt}
Let $\r$ be an $M\times N$ PPT state of rank $r<M+N-2$ and let
$r'=M+N-1-r$. Then

(i) $\ker\r$ contains subspaces $\ket{a}\ox W$ and $V\ox\ket{b}$
of dimension $r'$.

(ii) The subspaces in (i) can be chosen so that $\ket{a}\in V$
and $\ket{b}\in W$.
 \et
 \bpf
(i) By symmetry, it suffices to prove only the assertion that
$\ker\r$ contains a subspace $V\ox\ket{b}$ of dimension $r'$.
By Theorem \ref{thm:PPTMxNrank<M,N} we have $r\ge\max(M,N)$
and so $\min(M,N)\ge3$.
We can write $\r$ as $\r=C^\dag C$, where $C=[C_0~\cdots~C_{M-1}]$
and the $C_i$ are $r\times N$ matrices.
Since $\ker\r$ contains a product vector, we may assume that
$\ket{0,N-1}\in\ker\r$. Hence, we may assume that the $C_i$
have the form
 \begin{equation}
 \label{ea:2x1=1x2ppt}
C_0=\left[\begin{array}{cc}
    I_R & 0
 \\ 0 & 0 \end{array}\right]; \quad
C_i=\left[\begin{array}{cc}
    C_{i0} & C_{i1}
 \\ C_{i2} & C_{i3} \end{array}\right],~ i>0,
 \end{equation}
where the blocks $C_{i0}$ are $R\times R$. Since $\r^\G\ge0$ we must
have $C_{i1}=0$, $i>0$.
Let $m$ be the dimension of the matrix space spanned by the blocks
$C_{i3}$ and note that $m\ge1$.
We can now assume that the
blocks $C_{i3}$, $i=1,\ldots,m$, are linearly independent and
$C_{i3}=0$ for $i>m$.

We use the induction on $M+N$, and for fixed $M$ and $N$ the
induction on $r$, to prove the above assertion.
Let us assume that it holds for PPT states of rank less than $r$.
This is vacuously true when $r=\max(M,N)$.
If $r\ge m+N-1$ then the assertion follows from the observation that
$\{\ket{1},\ldots,\ket{m}\}^\perp\ox\ket{N-1}\subseteq\ker\r$
and $M-m\ge r'$.
Assume that $r<m+N-1$ and let us apply the induction hypothesis to
the PPT state $\s:=(C')^\dag C'$ where $C'=[C_{13}~\cdots~C_{m3}]$.
This state acts on a $(M-1)\ox(N-R)$ subsystem of our $M\ox N$
system, and we have $\rank\s\le r-R$ and $\rank\s_A=m$.
Since $\r_B>0$, and $\s_B$ is a principal submatrix of $\r_B$,
it follows that $\s_B>0$. In particular, $\rank\s_B=N-R$.
Thus $\s$ is an $m\times(N-R)$ PPT state of rank at most $r-R$.
Since $\rank\s\le r-R\le m+(N-R)-2$, we infer that there exists a
subspace $V'\ox\ket{b}\se\ker\s$ of dimension at least $m+N-1-r$.
(If $\rank\s=m+(N-R)-2$ we know this is true without using the
induction hypothesis.)
By applying an ILO on party B of the $(M-1)\ox(N-R)$ subsystem,
we may assume that $\ket{b}=\ket{N-1}$.
Since the sum $V''=V'+\{\ket{1},\ldots,\ket{m}\}^\perp$ is direct,
we have $\dim V''\ge(m+N-1-r)+(M-m)=r'$.
For any $r'$-dimensional subspace $V$ of $V''$, we have
$V\ox\ket{N-1}\subseteq\ker\r$ and the proof is completed.

(ii) By invoking (i) we can assume that $N-R\le r'$.
Then, because $\ket{0}_A\in V''$, we can choose $V$ so that
$\ket{0}_A\in V$.
Similarly, we can choose an $r'$-dimensional subspace $W$ contained
in the span of the basis vectors $\ket{R}_B,\ldots,\ket{N-1}_B$
such that $\ket{N-1}_B\in W$.
It remains to observe that $\ket{0}\ox W$ and $V\ox\ket{N-1}$
are contained in $\ker\r$.
 \epf

The analog of assertion (i) for bad $M\times N$ PPT states of rank
$r=M+N-2$ is not valid. A counter-example is the state $\r$ in
Example \ref{Besk-Fam} with $a=b=c=d=e=1$ and $f=g=0$. On one hand
we have $\ket{0}_A\ox W\subseteq\ker\r$ where $W$ is the span of
$\ket{2}_B$ and $\ket{3}_B$. On the other hand, by using Eq.
(\ref{JezgroRo}), it is not hard to show that $\ker\r$ contains no
two-dimensional subspace $V\ox\ket{y}_B$.

We conclude this section with another property of states whose kernel
contains only finitely many product vectors.

 \bpp
 \label{prop:NECESSARY=good}
Let $\r$ be an $M\times N$ state such that $|X_\r|<\infty$.
If $\rank\r^\G=M+N-2$ then $\r$ is a good PPT state of rank $M+N-2$.
 \epp
 \bpf
If $\ket{x,y}\in\ker\r^\G$, then
$\bra{x^*,y}\r\ket{x^*,y}=\bra{x,y}\r^\G\ket{x,y}=0$.
As $\r\ge0$, we have $\ket{x^*,y}\in\ker\r$.
We infer that $|X_{\r^\G}|<\infty$.
Let $\ket{\psi}$ be an eigenvector of $\r^\G$ with eigenvalue
$\lambda\ne0$ and let $H=\bC\ket{\psi}+\ker\r^\G$.
Since $\dim H>(M-1)(N-1)+1$, $H$ contains infinitely
many product vectors. Hence, there exists $\ket{\phi}\in\ker\r^\G$
such that $\ket{\psi}+\ket{\phi}=\ket{a,b}$ is a product vector.
Since $\r^\G\ket{\phi}=0$, we have
$\bra{\psi}\r^\G\ket{\psi}=\bra{a,b}\r^\G\ket{a,b}
=\bra{a^*,b}\r\ket{a^*,b}\ge0$.
It follows that $\lambda>0$. Hence $\r^\G\ge0$, i.e.,
$\r$ is a PPT state. By Theorem \ref{thm:delta} $\r^\G$ is good.
Thus we can apply Theorem \ref{thm:LMS2010(ii),ker=finitePRO} to
$\r^\G$ to complete the proof.
 \epf

\section{$M\times N$ PPT states of rank $N+1$}
\label{sec:N+1}

In this section we focus on part (ii) of Conjecture
\ref{conj:LMS2010highdimension}.
The Proposition \ref{prop:MxNrank(N+1)PPT} and
Theorems \ref{thm:MxNrank(N+1)reducible} and
\ref{thm:MxNrank(N+1)=3exclusive} describe  the structure of
the $M\times N$ PPTES $\r$ of rank $N+1$.
The main result of this section is
Theorem \ref{thm:MxNrank(N+1)=reducible,M>3} where we show that,
for $M,N>3$ any $M\times N$ PPTES $\r$ of rank $N+1$ is reducible.
Hence, such states cannot be extreme. From this result we deduce
that $\cE_r^{M,N}=\emptyset$ if $\min(M,N)=3,4$ and $1<r<M+N-2$.
Theorem \ref{thm:3xN-PPTESbirank} will be used in the next section
for the construction of extreme states.

The following proposition is an analog of
\cite[Proposition 18]{cd11JPA}. Recall that the direct sum of
two states was introduced in Definition \ref{def:reducibleirreducible}.
 \bpp
 \label{prop:MxNrank(N+1),commonkernel,PPT}
Let $\r$ be an $M\times N$ PPT state and let $\ket{a}\in\cH_A$ be
such that $\rank\bra{a}\r\ket{a}=1$. Then $\r=\r_1\oplus_A\r_2$
where $\r_1$ is a pure product state. If $\r$ is entangled and
$\rank\r=N+1$, then $\r=\r_1\oplus\r_2$.
 \epp
 \bpf
By using Eq. (\ref{MxN-State}) we have $\r=C^\dag C$, where
$C=[C_0~C_1~\cdots~C_{M-1}]$ and the $C_i$ are matrices of size $R
\times N$, $R=\rank\r$. By choosing suitable bases we can assume
that $\ket{a}=\ket{0}_A$ and $\ker\bra{a}\r\ket{a}=\ket{0}_B^\perp$.
Consequently, only the first column of $C_0$ is nonzero. By
replacing $C$ with $UC$ where $U$ is unitary, we may also assume
that the first column of $C_0$ is 0 except for its first entry which
is nonzero. By rescaling $C_0$, we may assume that this entry is 1.
Since $\r^\G\ge0$ and only the first entry of $C_0^\dag C_0$ is
nonzero, we infer that the first row of $C_i$, $i>0$, is 0 except
possibly its first entry. By subtracting suitable multiples of $C_0$
from the $C_i$, $i>0$, we may assume that the first rows of these
$C_i$ are 0. It is now easy to check that $\r=\proj{00}\oplus_A\r_2$
where $\r_2={C'}^\dag C'$ and $C'$ is the submatrix of
$[C_1~C_2~\cdots~C_{M-1}]$ obtained by deleting the first row. The
first assertion is proved.

Now assume that $\r$ is entangled and that $\rank\r=N+1$. Then
$\rank\r_2=N$ and $\r_2$ must be entangled by Lemma
\ref{le:reducible=SUMirreducible,SEP,PPT}. Clearly, we have
$\rank(\r_2)_B\le N$. On the other hand, since
$\r_B=\proj{0}+(\r_2)_B$ we have $\rank(\r_2)_B\ge\rank\r_B-1=N-1$.
Hence, Proposition \ref{prop:PPTMxNrankN} implies that
$\rank(\r_2)_B=N-1$. It follows from the first assertion that
$\rank(\r_2)_A=M-1$, and so the second assertion is proved.
 \epf

We start by assuming that the range of a PPT state $\r$ contains a product vector in which case it is relatively easy to describe
the structure of $\r$.
 \bpp
 \label{prop:MxNrank(N+1)PPT}
$(M,N>2)$ Let $\r$ be an $M\times N$ PPT state of rank $N+1$ such
that $\cR(\r)$ contains at least one product vector.
If $\r$ is B-irreducible, then $\r$ is a sum of $N+1$
pure product states.
Otherwise, $\r=\r_1\oplus_B\r_2$ where $\r_1$
is a pure product state.
 \epp
 \bpf
In order to prove the second assertion, let us assume that
$\r=\r'\oplus_B\r''$.
Since $\rank\r'+\rank\r''=N+1$ and $\rank\r'_B+\rank\r''_B=N$,
we may assume that $\rank\r'=\rank\r'_B$.
Hence, we can apply Proposition \ref{prop:PPTMxNrankN} to $\r'$.
As the sum in this proposition is necessarily B-direct, the
second assertion is proved.

From now on we assume that $\r$ is B-irreducible.
By Proposition \ref{prop:MxNrank(N+1),commonkernel,PPT}, we have
$\rank\bra{b}\r\ket{b}\ge2$ for all nonzero $\ket{b}\in\cH_B$.
Using Eq. (\ref{MxN-State}), we have $\r=C^\dag C$ where
$C=[C_0~C_1~\cdots~C_{M-1}]$ and the $C_i$ are $(N+1)\times N$
matrices.

Assume that there is an $\ket{a}\in\cH_A$ such that
$\rank\bra{a}\r\ket{a}=1$.
By Proposition \ref{prop:MxNrank(N+1),commonkernel,PPT}
$\r$ is an A-direct sum of a pure product state and an
$(M-1)\times P$ state $\s$ of rank $N$.
Since $\r$ is B-irreducible, we must have $P=N$.
Hence, the first assertion holds in this case by
Proposition \ref{prop:PPTMxNrankN}.
Thus we may assume that $\rank\bra{a}\r\ket{a}\ge2$ for all nonzero
vectors $\ket{a}\in\cH_A$.
In particular, $\rank C_i\ge2$ for each $i$.

By the hypothesis, we may assume that the first row of $C$ corresponds
to the product vector in $\cR(\r)$. By performing an ILO on
system A, we may also assume that the first row of each $C_i$,
$i>0$, is 0. The state
$\s:=[C_1~\cdots~C_{M-1}]^\dag[C_1~\cdots~C_{M-1}]$ is PPT and
$\s_B=\sum_{i>0} C_i^\dag C_i$.
If $\s_B\ket{b}=0$ for some $\ket{b}\ne0$, then $C_i\ket{b}=0$
for $i>0$ and so $\ket{0}^\perp\ox\ket{b}\subseteq\ker\r$.
This contradicts our assumption on the rank of $\bra{b}\r\ket{b}$.
We conclude that $\rank\s_B=N$.
Since $\s$ is PPT and $\rank\s \le N$, Theorem \ref{thm:PPTMxNrank<M,N}
implies that $\rank\s=N$ and $M\le N$.

By dropping the first row of $C_i$, we obtain the $N\times N$ matrix
$C'_i$, $i=0,1,\ldots,M-1$. By applying Proposition
\ref{prop:PPTMxNrankN} and \cite[Proposition 6]{cd11JPA} to the
state $\s$, we may assume that the matrices $C'_i$, $i>0$, are
diagonal. Since $\rank\s=N$, we may also assume that $C'_1=I_N$. By
simultaneously permuting the diagonal entries (if necessary) we may
assume that
 \bea \label{jed:C'-blokovi}
 C'_i=\l_{i1} I_{l_1}\oplus\cdots\oplus \l_{ik} I_{l_k},
 \quad i>0; \quad l_1 + \cdots + l_k=N,
 \eea
and that whenever $r\ne s$ there exists an $i>1$ such that
$\l_{ir}\ne\l_{is}$. (Note that all $\l_{1r}=1$.) Since the $C_i$
are linearly independent, each set $\{\l_{ir}:r=1,\ldots,k\}$,
$i>1$, must have at least two elements. In particular, we have
$k\ge2$. The local transformations that we used to transform the
$C'_i$, $i>0$, to this special form, can be performed on the entire
matrices $C_i$, $i>0$. In order to transform simultaneously the
state $\r$, we have to perform the same local B-transformations on
$C_0$ as well as to multiply it by the same unitary matrices on the
left hand side. The first rows of the $C_i$, $i>0$, are not affected
by any of these transformations and will remain 0.

We partition the matrix $C'_0=[A_{ij}]_{i,j=1}^k$ with $A_{ii}$
square of order $l_i$. We claim that $A_{rs}=0$ for $r\ne s$. To
prove this claim, recall that there exists an index $i>1$ such that
$\l_{ir}\ne\l_{is}$. We may assume temporarily that $\l_{is}=0$.
(Just replace $C_i$ with $C_i-\l_{is} C_1$.) Then the $s$th diagonal
block of order $l_s$ in $C_i^\dag C_i$ is 0. Since
 \bea
\left[\begin{array}{cc} C_0^\dag C_0 & C_0^\dag C_i \\ C_i^\dag C_0
& C_i^\dag C_i
\end{array}\right]^\G
=\left[\begin{array}{cc} C_0^\dag C_0 & C_i^\dag C_0 \\ C_0^\dag C_i
& C_i^\dag C_i
\end{array}\right]\ge0,
 \eea
we deduce that the $s$th block-row of $C_0^\dag C_i$ must vanish. In
particular, $\l_{ir}A_{rs}^\dag=0$. As $\l_{ir}\ne\l_{is}=0$, our
claim is proved.

Hence, we have $C'_0=B_1\oplus\cdots\oplus B_k$ with $B_i=A_{ii}$
square of order $l_i$. Let $U_i$ be a unitary matrix such that
$U_iB_iU_i^\dag$ is upper triangular and let
$U=U_1\oplus\cdots\oplus U_k$. Note that the transformation $C_i\to
([1]\oplus U)C_i U^\dag$ leaves the matrices $C_i$, $i>0$,
unchanged. Thus we may assume that all $B_i$ are upper triangular.
The first row of $C_0$ consists of the vectors $w_1,\ldots,w_k$ of
lengths $l_1,\ldots,l_k$, respectively. Let $\m_i$ and $\nu_i$ be
the first entries of $w_i$ and $B_i$, respectively.

If some $\m_i$ is 0, say $\m_1=0$, then by subtracting from $C_i$,
$i\ne1$, a suitable scalar multiple of $C_1$, we may assume that the
first columns of these $C_i$ are 0.
This contradicts our assumption on the rank of $\bra{b}\r\ket{b}$.
Hence, all $\m_i\ne0$.

We claim that, for any
$s\in\{1,\ldots,k\}$, the matrix $B_s$ is diagonal. To prove this
claim, let us choose an $r\in\{1,\ldots,k\}$ such that $r\ne s$. Let
us also fix an index $i>1$ such that $\l_{ir}\ne\l_{is}$. (Recall
that such $i$ exists.) Since $[C_0~C_1~C_i]^\dg[C_0~C_1~C_i]$ is a
PPT state, so is $[\hat{C}_0,\hat{C}_i]^\dg[\hat{C}_0,\hat{C}_i]$
where $\hat{C}_0=C_0-\nu_r C_1$ and $\hat{C}_i=C_i-\l_{ir} C_1$.
Since $\m_r$ is the only nonzero entry in the
$(l_1+\cdots+l_{r-1}+1)$th column of $\hat{C}_0$, and the
corresponding column of $\hat{C}_i$ is 0, we may assume that $\m_r$
is the only nonzero entry in the first row of $\hat{C}_0$. It
follows that the state
 \bea \label{StanjeB}
 [B_s-\nu_r I_{l_s}~(\l_{is}-\l_{ir})I_{l_s}]^\dag
 [B_s-\nu_r I_{l_s}~(\l_{is}-\l_{ir})I_{l_s}]
 \eea
is PPT. Since $\l_{ir}\ne\l_{is}$, the state (\ref{StanjeB}) is
a $2\times l_s$ state of rank $l_s$.
Hence it is separable and, by \cite[Proposition 6]{cd11JPA},
$B_s$ is a normal matrix. Since it is also upper triangular,
it must be diagonal. Hence, our claim is proved.

It follows that $\r$ is a sum of $N+1$ pure product states, which
completes the proof of the first assertion.
 \epf

\begin{example} \label{eg:3x3birank(4,5)}
$(M=N=3)$ {\rm As Proposition \ref{prop:MxNrank(N+1)PPT} suggests,
a separable $M\times N$ state of rank $N+1$ may fail to be the sum of
$N+1$ pure product states. Indeed, the $3\times3$ separable state
$\r=2\sum^2_{i=0}\proj{ii}+(\ket{01}+\ket{10})(\bra{01}+\bra{10})$
has rank four. As $\r^\G$ has rank five, $\r$ is not a sum of four
pure product states. \hfill $\square$ }
\end{example}

\begin{example} \label{eg:3x4rank5,onlyBdirecsum}
$(M=3,N=4)$ {\rm As Proposition \ref{prop:MxNrank(N+1)PPT}
suggests, an $M\times N$ PPTES of rank $N+1$ may be A-irreducible.
As an example we can take the $3\times4$ state $\r=\proj{00}\op_B\s$
of rank five, where $\s$ is a $3\times3$ PPTES of rank four.
Suppose $\r=\r_1\op_A\r_2$. Then we have $\rank(\r_i)_A\le2$ and
$\rank\r_i\le4$. Thus both $\r_1$ and $\r_2$ are separable,
and so is $\r$. We have a contradiction. \hfill $\square$ }
\end{example}

For any $2\times N$ state $\r$ of rank $N+1$, $\cR(\r)$ contains
infinitely many product vectors, see Eq. (\ref{jed:delta}). The
first example $\r$ of $2\times4$ PPTES was constructed in
\cite{woronowicz76} and \cite[Eq. (32)]{horodecki97}. This state
has rank five and so $\cR(\r)$ contains infinitely many product
vectors. Moreover, we claim that $\r$ is irreducible. To prove
this claim, assume that $\r$ is reducible. Then necessarily
$\r=\r_1\oplus_B\r_2$, and by Lemma
\ref{le:reducible=SUMirreducible,SEP,PPT} $\r_1$ and $\r_2$ are PPT.
Since their B-local ranks are at most three, they are separable.
This is a contradiction and the claim is proved. Thus Proposition
\ref{prop:MxNrank(N+1)PPT} does not extend to the case $M=2$.

We can now characterize the reducible $M\times N$ PPTES of rank
$N+1$.

 \bt
 \label{thm:MxNrank(N+1)reducible}
$(M,N>2)$ For an $M\times N$ PPTES $\r$ of rank $N+1$,
the following are equivalent to each other

(i) $\r$ is reducible;

(ii) $\cR(\r)$ contains at least one product vector;

(iii) $\r=\r_1\oplus_B\r_2$, where $\r_1$ is a pure product state.
 \et
 \bpf
(i) $\ra$ (ii). Assume $\r=\r'+\r''$ is an A or B-direct sum. By
Theorem \ref{thm:PPTMxNrank<M,N} we have $\rank\r'\ge\rank\r'_B$
and $\rank\r''\ge\rank\r''_B$.
Since $\rank\r'+\rank\r''=\rank\r=N+1$, we have
$N+1\ge\rank\r'_B+\rank\r''_B\ge\rank\r_B=N$.
Therefore, say, $\rank\r'= \rank\r'_B$. Hence $\r'$ is
separable by Proposition \ref{prop:PPTMxNrankN}, and (ii) holds.

(ii) $\ra$ (iii) follows from Proposition \ref{prop:MxNrank(N+1)PPT}
because $\r$ is entangled.

(iii) $\ra$ (i) is trivial.
 \epf

Using these results, we now prove the main result of this section.

\bt $(M,N>3)$
 \label{thm:MxNrank(N+1)=reducible,M>3}
If $\r$ is an $M\times N$ PPTES of rank $N+1$ then
$\r=\r_1\oplus_B\r_2$, where $\r_1$ is a pure product state.
Consequently, $\cE^{M,N}_{N+1}=\emptyset$.
 \et
 \bpf
Let $\r$ be an $M\times N$ PPTES of rank $N+1$.
Suppose that the assertion is false.
Then, by Theorem \ref{thm:MxNrank(N+1)reducible},
$\r$ is irreducible and $\cR(\r)$ is a CES.
For any $\ket{a}\in\cH_A$ let $r_a$ be the rank of the linear
operator $\bra{a}\r\ket{a}$.
Since $\dim\ker\r=MN-N-1>(M-1)(N-1)+1$, $\ker\r$ contains infinitely
many product vectors. If $\ket{a,b}\in\ker\r$ is a product vector
then $\bra{a}\r\ket{a}$ kills the vector $\ket{b}$, and so $r_a<N$.
Let $R$ be the maximum of $r_a$ taken over all $\ket{a}\in\cH_A$
such that $r_a<N$. Thus $R<N$. Without any loss of generality we
may assume that $\bra{0}_A\r\ket{0}_A$ has rank $R$.

We can write $\r$ as in Eq. (\ref{MxN-State}). Thus $\r=C^\dag C$
where $C=[C_0~\cdots~C_{M-1}]$ and the blocks $C_i$ are $(N+1)\times
N$ matrices. By Proposition
\ref{prop:MxNrank(N+1),commonkernel,PPT}, we have $r_a>1$ for all
nonzero vectors $\ket{a}\in\cH_A$. In particular, $\rank C_i\ge2$
for each $i$. Consequently, we may assume that
 \begin{equation}
C_0=\left[\begin{array}{cc}
    I_R & 0
 \\ 0 & 0 \end{array}\right]; \quad
C_i=\left[\begin{array}{cc}
    C_{i0} & C_{i1}
 \\ C_{i2} & C_{i3} \end{array}\right],~ i>0,
 \end{equation}
where the $C_{i0}$ are $R\times R$ matrices. Since $\r^\G\ge0$,
all $C_{i1}=0$.

The state $\s={C'}^\dag C'$, where $C'=[C_{13}~\cdots~C_{M-1,3}]$,
is a PPT state of rank $\le N-R+1$ which acts on a $(M-1)\ox(N-R)$
subsystem of our $M\ox N$ system. Since $\r_B>0$ and $\s_B$ is its
principal submatrix, we have $\rank\s_B=N-R$.
By using Theorem \ref{thm:PPTMxNrank<M,N}, we deduce that the rank
of $\s$ must be either $N-R$ or $N-R+1$. Assume that this rank is
$N-R$. Then, by Proposition \ref{prop:PPTMxNrankN}, $\s$ is a sum
of $N-R$ pure product states. Consequently, we may assume that the
blocks $C_{i3}$ are diagonal matrices (with the zero last row).
Moreover, we can assume that the first entry of $C_{i3}$ is 1
for $i=1$ and 0 for $i>1$. Since $\r^\G\ge0$,
the first row of $C_{i2}$, $i>1$, must be 0. Thus the nonzero
entries of the $(R+1)$st row of $C$ occur only inside the block $C_1$.
This means that $\cR(\r)$ contains a product vector, which gives
us a contradiction.

We conclude that $\s$ must have rank $N-R+1$, and so $m:=\rank\s_A$
is in the range $1<m<M$.
Hence, we may assume that $C_{i3}=0$ for $i>m$.
Consequently, the matrices $C_{i3}$, $1\le i\le m$ are linearly
independent. Moreover, by using the definition of $R$, we know that
any nontrivial linear combination of these $m$ matrices must
have full rank, $N-R$.

Assume now that $m>2$. We can consider the state $\s$ as acting on
the Hilbert space $\cR(\s_A)\ox\cR(\s_B)$ of dimension $m(N-R)$.
Then its kernel has the dimension $(m-1)(N-R)-1$ which is bigger
than $(m-1)(N-R-1)$. Therefore this kernel contains a product
vector. Equivalently (see Eq. (\ref{JezgroRo})), there exist scalars
$\xi_i$, $i=1,\ldots,m$, not all 0, such that the matrix
$\sum_{i=1}^m \xi_i C_{i3}$ has rank less than $m$. Thus we have a
contradiction.

Consequently, we must have $m=2$. Since any nontrivial linear
combination of $C_{13}$ and $C_{23}$ has rank $N-R$,
the matrix $[C_{13}~C_{23}]$ must have rank $N-R+1$.
For $i>2$ we have $C_{i3}=0$ and since $\r^\G\ge0$,
it follows that $C_{i2}^\dag C_{13}=C_{i2}^\dag C_{23}=0$.
Hence, we have $C_{i2}=0$ for $i>2$.

The state $\tau:={C''}^\dag C''$, where
$C''=[I_R~C_{30}~\cdots~C_{M-1,0}]$, is a PPT state of rank $R$.
Note that $\rank\tau_A\le M-2$ and $\rank\tau_B=R$.
By Proposition \ref{prop:PPTMxNrankN} we can assume that the
matrices $C_{i0}$, $i>2$, are diagonal.
By simultaneously permuting their diagonal entries (if necessary)
we may assume that
 \bea
 C_{i0}=\l_{i1} I_{l_1}\oplus\cdots\oplus \l_{ik} I_{l_k},
 \quad i>2; \quad l_1 + \cdots + l_k=R,
 \eea
and that whenever $r\ne s$ there exists an $i$ such that
$\lambda_{ir}\ne\lambda_{is}$.
Since the blocks $C_{ij}=0$ when $i>2$ and $j\ne0$ and
$\rank\r_A=M\ge4$, we must have $k>1$.

As in the proof of Proposition \ref{prop:MxNrank(N+1)PPT}, we can
show that the matrices $C_{10}$ and $C_{20}$ are direct sums
 \bea
C_{10}= E_{1}\oplus\cdots\oplus E_{k},\quad
C_{20}= F_{1}\oplus\cdots\oplus F_{k},
 \eea
where $E_{i}$ and $F_{i}$ are square blocks of size $l_i$,
and we may assume the $E_{i}$ are lower triangular.

Let us write
 \bea
C_{i2}=\left[
          \begin{array}{c}
            C_{i21} \\
            C_{i22} \\
          \end{array}
        \right],\quad i>0;\quad
C_{i3}=\left[
          \begin{array}{c}
            C_{i31} \\
            C_{i32} \\
          \end{array}
        \right],\quad i=1,2;
 \eea
where $C_{i22}$ and $C_{i32}$ are row-vectors. By multiplying $C$ on
the left hand side by a unitary matrix $I_R\oplus U$, we may assume
that $C_{132}=0$. Since $C_{13}$ has rank $N-R$, the block $C_{131}$
is an invertible matrix. Consequently, we may assume that
$C_{121}=0$. We split the row-vector $C_{122}$ into $k$ pieces
$w_1,\ldots,w_k$ of lengths $l_1,\ldots,l_k$, respectively. To
summarize, the matrices $C_j$, $j>0$, have the form:
 \begin{equation*}
C_1=\left[\begin{array}{cc}
    \left[\begin{array}{ccc}
       E_1 & & \\
       & \ddots & \\
       & & E_k \\
      \end{array}\right]
    & 0
 \\ 0 & C_{131}
 \\
 \left[
 \begin{array}{ccc} w_1,\ldots,w_k \end{array} \right]
    & 0
 \end{array}\right],
 \quad
C_2=\left[\begin{array}{cc}
     \left[\begin{array}{ccc}
       F_1 & & \\
       & \ddots & \\
       & & F_k \\
      \end{array}\right]
    & 0
 \\ C_{221} & C_{231}
 \\ C_{222} & C_{232} \end{array}\right];
 \quad
C_j=\left[\begin{array}{cc}
    \left[\begin{array}{ccc}
       \l_{j1} I_{l_1} & & \\
       & \ddots & \\
       & & \l_{jk} I_{l_k} \\
      \end{array}\right]
    & 0
 \\ 0 & 0
 \\ 0 & 0 \end{array}\right],\quad j>2.
 \end{equation*}
Since $\cR(\r)$ is a CES, each $l_i>1$ and at least one $w_i\ne0$.
As we can simultaneously permute the first $k$ diagonal blocks
of the matrices $C_j$, we may assume that $w_1\ne0$.
Let $w_1=(a_1,\ldots,a_n,0,\ldots,0)$ where $a_n\ne0$ and let us
partition
 \bea
E_1= \left[\begin{array}{cc}
             E_{10} & 0 \\
             E_{12} & E_{13} \\
           \end{array}\right],
 \eea
where $E_{10}$ is of size $n\times n$.

If $n<l_1$ then the state
$[I_{l_1-n}~E_{13}]^\dg[I_{l_1-n}~E_{13}]$ is PPT and so the matrix
$E_{13}$ must be normal.
Since $E_{13}$ is also lower triangular, it must be a diagonal matrix.
By using this fact and the observation that the state
$[C_0~C_1]^\dg[C_0~C_1]$ is PPT, one can easily show that $E_{12}=0$.
We conclude that except for $a_n$ and the last entry of $E_{10}$ all
other entries of the $n$th column of $C_1$ are 0.
This is trivally true also in the case $n=l_1$.
By subtracting from $C_1$ a scalar multiple of $C_0$,
we may assume that the last entry of $E_{10}$ is 0.
Now $a_n$ is the only nonzero entry in the $n$th column of $C_1$.

We can choose an index $i>2$ such that $\l_{i1}\ne\l_{ik}$.
By replacing temporarily $C_i$ with $C_i-\lambda_{i1}C_0$, the $n$th column of $C_i$ becomes 0.
It follows easily that the state
$[E_k,(\l_{ik}-\l_{i1})I_{l_k}]^\dg[E_k,(\l_{ik}-\l_{i1})I_{l_k}]$
is a PPT state of rank $l_k$.
Since its B-local rank is also $l_k$, the matrix $E_k$ must be
normal. As it is also lower triangular, it must be a diagonal matrix.
We can further assume that
 \bea
 E_k = \m_{1} I_{n_1}\oplus\cdots\oplus \m_{s} I_{n_{s}}, \quad
 F_k = G_{n_1}\oplus\cdots\oplus G_{n_{s}};
 \quad n_1 + \cdots + n_{s} = l_k,
 \eea
with each $G_j$ upper triangular of order $n_j$.
Then the $R$th row of $C$ shows that $\cR(\r)$ contains a product
vector.

This contradicts Theorem \ref{thm:MxNrank(N+1)reducible}, and so
the proof is completed.
 \epf

As a consequence, we obtain a link between the good and extreme
states.

 \bpp
$(N\ge M=3,4)$ \label{pp:TeoremaSep}
Let $\r$ be a good $M\times N$ PPT state of rank $M+N-2$.

(i) If $\r$ is entangled then it is strongly extreme.

(ii) If $\cR(\r)$ contains a product vector, then $\r$ is
separable.
 \epp
 \bpf
(i) follows from Theorem \ref{thm:MxNrank(N+1)=reducible,M>3}
and parts (iii) and (v) of Theorem
\ref{thm:LMS2010(ii),ker=finitePRO}.

(ii) If $M=3$, it follows from Theorem
\ref{thm:LMS2010(ii),ker=finitePRO} (i) that a good $\r$ is
irreducible. Then $\r$ is separable by Proposition
\ref{prop:MxNrank(N+1)PPT}.

Now let $M=4$, and assume that $\r$ is entangled.
Since $\r$ is good, part (iii) of
Theorem \ref{thm:LMS2010(ii),ker=finitePRO} shows that $\r$ is not
extreme and so we have $\r=\r_1+\r_2$ with $\r_1$ a PPTES.
It follows from part (iv) of the same theorem that $\r_1$ is a
$4\times N$ state. For the same reason and
Theorems \ref{thm:MxNrank(N+1)reducible}
and \ref{thm:MxNrank(N+1)=reducible,M>3}, we have $\rank\r_1=N+2$.
By  Theorem \ref{thm:LMS2010(ii),ker=finitePRO} (iii),
$\r_2\propto\r_1$ and $\r$ is extreme. This is a contradiction,
and so we conclude that $\r$ must be separable.
 \epf

When $M=N=3$ the hypothesis that $\r$ is good can be removed, see
Theorem \ref{thm:3x3PPTstates} (i).

The following theorem, which extends parts (i) and (ii) of
Theorem \ref{thm:3x3PPTstates} to $M\ox N$ systems, follows easily
from Theorems \ref{thm:MxNrank(N+1)reducible} and
\ref{thm:MxNrank(N+1)=reducible,M>3}.

 \bt
 \label{thm:MxNrank(N+1)=3exclusive}
$(M,N>2)$ Let $\r$ be an $M\times N$ PPT state of rank $N+1$.
Then $\r=\r_1\oplus_B\cdots\oplus_B\r_k\oplus_B\s$,
where $\r_i$ are pure product states, $k\ge0$, and $\s$ is
B-irreducible.
If $\r$ is entangled, then $\rank\s_A=2$ or $3$.
 \et
For $M=3$ see Example \ref{Besk-Niz}.

The next result shows that certain states whose range is contained
in the range of an irreducible $M\times N$ PPTES of rank $N+1$ are
also $M\times N$ PPTES of rank $N+1$.

 \bt
 \label{thm:MxNrank(N+1),R>rank rhoB}
$(M,N>2)$ Let $\r$ be an irreducible $M\times N$ PPTES of rank
$N+1$.

(i) Any state $\r_1$ with $\cR(\r_1)\subseteq\cR(\r)$ and
$\rank\r_1>\rank(\r_1)_B$ is an $M\times N$ state of rank $N+1$.

(ii) Any PPT state $\r_1$ with $\cR(\r_1)\subseteq\cR(\r)$ is an
irreducible $M\times N$ PPTES of rank $N+1$.
 \et
 \bpf
(i) Without any loss of generality, we may assume that
$\r_2:=\r-\r_1\ge0$. By Theorem \ref{thm:MxNrank(N+1)reducible},
$\cR(\r)$ is a CES, and so both $\r_1$ and $\r_2$ must be entangled.
Let $R_i=\rank\r_i$ and $r_i=\rank(\r_i)_B$, $i=1,2$. By the
hypothesis we have $R_1>r_1$.

Assume that $r_1<N$. By using Eq. (\ref{MxN-State}), we may assume
that $\r=C^\dag C$, where $C=[C_0~C_1~\cdots~C_{M-1}]$ and
 \begin{equation} \label{Blok-Mat}
C_i=\left[\begin{array}{cc}
    C_{i0} & C_{i1}
 \\ 0 & C_{i2} \end{array}\right]
 \end{equation}
are matrices of size $(R_1+R_2)\times N$, the blocks $C_{i0}$ are of
size $n\times(N-r_1)$, $n\le R_2$, and the matrix
$[C_{00}~C_{10}~\cdots~C_{M-1,0}]$ has rank $n$. The bottom $R_1$
[top $R_2$] rows of $C$ represent $\r_1$ $[\r_2]$. Since
$\rank\r=N+1$, we can choose a unitary matrix $U$ such that the
bottom $R_1+R_2-N-1$ rows of $U[C_{02}~C_{12}~\cdots~C_{M-1,2}]$ are
0. Then the last $R_1+R_2-N-1$ rows of $(I_n\oplus U)C$ are 0, and
by dropping them, we may assume that in Eq. (\ref{Blok-Mat}) the
$C_i$ are of size $(N+1)\times N$ and, as before, the blocks
$C_{i0}$ have size $n\times (N-r_1)$. Since
 \bea
C_i^\dag C_j=\left[\begin{array}{cc}
  C_{i0}^\dag C_{j0} & C_{i0}^\dag C_{j1} \\
  C_{i1}^\dag C_{j0} & C_{i1}^\dag C_{j1}+C_{i2}^\dag C_{j2}
\end{array}\right],
 \eea
it follows immediately that the state $\s:=[C_{i0}^\dag
C_{j0}]_{i,j=0}^{M-1}$ is PPT and that $\s_B = \sum_i C_{i0}^\dag
C_{i0}$ has rank $N-r_1$. By Theorem \ref{thm:PPTMxNrank<M,N}, we
have $N+1-R_1\ge n=\rank\s\ge\rank\s_B=N-r_1$. As $R_1>r_1$, we must
have $R_1=r_1+1$ and $\rank\s=N-r_1$. By Proposition
\ref{prop:PPTMxNrankN}, $\s$ is separable and is the sum of $N-r_1$
pure product states. Consequently, as mentioned in Sect.
\ref{sec:preliminary}, we may assume that the blocks $C_{i0}$ are
diagonal matrices. We may also assume that the first entry of $C_0$
is not 0. By subtracting a scalar multiple of $C_0$ from the $C_i$,
$i>0$, we may assume that the first column of $C_i$ is 0. Then
Proposition \ref{prop:MxNrank(N+1),commonkernel,PPT} implies that
$\r$ is reducible, which is a contradiction.

Thus we must have $r_1=N$, and so $R_1=N+1$. Since
$\cR(\r_1)=\cR(\r)$, Lemma \ref{le:Range} implies that
$\rank(\r_1)_A=\rank\r_A=M$. Thus (i) holds.

(ii) By Theorem \ref{thm:MxNrank(N+1)reducible}, $\cR(\r)$ is a CES.
Hence, $\r_1$ is a PPTES. Theorem \ref{thm:PPTMxNrank<M,N} and
Proposition \ref{prop:PPTMxNrankN} imply that
$\rank\r_1>\rank(\r_1)_B$. By part (i), $\r_1$ is an $M\times N$
state of rank $N+1$. Finally, Theorem
\ref{thm:MxNrank(N+1)reducible} implies that $\r_1$ is irreducible.
 \epf

 \bt \label{thm:3xN-PPTESbirank}
Any irreducible $3\times N$ PPTES of birank $(N+1,N+1)$ is extreme.
 \et
 \bpf
Suppose there exists a counter-example, say $\r$. Since $\r$ is
irreducible, $\r^\G$ is also irreducible by Lemma
\ref{le:RHOreducible=RHOPTreducible}. Since $\r$ is not extreme,
$\r=\r_1+\r_2$ where $\r_1$ and $\r_2$ are non-parallel PPT states.
By Theorem \ref{thm:MxNrank(N+1),R>rank rhoB}, $\r_1$ and $\r_2$ are
$M\times N$ PPTES of rank $N+1$. Consequently, $\r_1,\r_2$ and $\r$
have the same range, and the same is true for $\r_1^\G,\r_2^\G$ and
$\r^\G$. Consider the Hermitian matrix $\s(t)=\r_2-t\r_1$ depending
on the real parameter $t$. Both $\s(t)$ and $\s(t)^\G$ are positive
semidefinite for $t\le0$ and indefinite for
$t=t_0:=\tr(\r_2)/\tr(\r_1)$. Hence, there exists a unique
$t_1\in(0,t_0)$ such that both matrices $\s(t)$ and $\s(t)^\G$ are
positive semidefinite and have rank $N+1$ for $0\le t<t_1$, while
at least one of them has rank $\le N$ for $t=t_1$. Since
$\r=(1+t_1)\r_1+\s(t_1)$ and $\r^\G=(1+t_1)\r_1^\G+\s(t_1)^\G$,
Theorem \ref{thm:MxNrank(N+1),R>rank rhoB} gives a contradiction.
 \epf

\section{Examples of $M\times N$ PPT states of rank $M+N-2$}
\label{sec:example}

If $\r$ is an $M\times N$ PPT state of rank $M+N-2$, then according
to Theorem \ref{thm:delta} there are two possibilities:
$\r$ is good in which case $\ker\r$ contains exactly $\d$ product vectors or $\r$ is bad in which case we know that $\ker\r$
contains a 2-dimensional subspace $V\ox W$.
Both cases occur even when $\r$ is a PPTES, and we will construct a variety of examples.
They are discussed in subsections A and B, respectively.
It follows immediately from Theorem
\ref{thm:LMS2010(ii),ker=finitePRO} that the states in the good case,
namely Examples \ref{Kon-Mnogo}, \ref{Besk-Fam,KERNEL=finite} and
\ref{Besk-Fam,KERNEL=finite,M=3,Nge4}, are strongly extreme.

\subsection{Good case: finitely many product vectors in the kernel}

Since we assume that $M,N>2$, the smallest case is $M=N=3$.
Let $M=N=3$ and let $\r$ be a $3\times3$ PPTES of rank four.
It is well-known that $\ker\r$ contains exactly six product
vectors. Hence, $\r$ is a good state by Proposition
\ref{prop:DobroStanje}.

Assuming that $M\le N$, the next case is $M=3$, $N=4$.
The state $\r$ of Example \ref{Kon-Mnogo} is extracted from the family {\bf GenTiles2} of UPB constructed in \cite{DiV03}. In this example, there are exactly ten product vectors in $\ker\r$
(which are not in general position). Next in Example
\ref{Besk-Fam,KERNEL=finite}, we shall construct a $3\times4$
extreme PPTES of rank five, whose kernel contains also exactly
ten product vectors. However, these product vectors are in
general position. This is the only known example of this kind.
At the end of this subsection, in Example
\ref{Besk-Fam,KERNEL=finite,M=3,Nge4}, we shall construct a
$3\times N$ extreme state of rank $N+1$ whose kernel contains
exactly $N(N+1)/2$ product vectors.

\begin{example} \label{Kon-Mnogo} {\rm ($M=3,N=4$)
Consider the 7-dimensional subspace $K$ of the space of complex
$3\times4$ matrices (identified with $\cH$):
 \bea
 \left[ \begin{array}{cccc}
\x_1+\x_7 & \x_4+\x_7 & -\x_3+\x_7 & -\x_4+\x_7 \\
-\x_1+\x_7 & \x_2+\x_7 & \x_5+\x_7 & -\x_5+\x_7 \\
\x_6+\x_7 & -\x_2+\x_7 & \x_3+\x_7 & -\x_6+\x_7
\end{array} \right] = \sum_{i=1}^7 \x_i W_i .
 \eea

The $W_i$ form an orthogonal (non-normalized) basis of $K$ and each
of them has rank one. Each of the matrices
\begin{eqnarray}
W_8 &=& 15(-W_1+W_3+W_5+W_6)-5W_4+3W_7, \\
W_9 &=& 15(W_1-W_2+W_4+W_6)-5W_5+3W_7, \\
W_{10} &=& 15(W_2-W_3+W_4+W_5)-5W_6+3W_7,
\end{eqnarray}
also has rank one. The orthogonal projector, $\r$, onto $K^\perp$ is
a $3\times 4$ PPT state of rank five. It is entangled because
$K^\perp$ is a CES. It is not hard to verify that $\ker\r=K$
contains only 10 matrices of rank one, namely the $W_i$,
$i=1,\ldots,10$.

Note that the 10 product vectors in $\ker\r$ are not in general
position. Indeed, if we write $W_i=\ket{a_i}\ox\ket{b_i}$ for each
$i$, then the $\ket{a_i}$ with $i=1,2,3$ are linearly dependent (and
the same is true for the $\ket{b_j}$ with $j=2,3,4,5$). \hfill
$\square$ }
\end{example}

We would like to construct examples of PPTES, $\r$, of rank $M+N-2$
such that $\ker\r$ contains exactly $\d$ product vectors, and
moreover these product vectors are in general position. An example
will be given later (see Example \ref{Besk-Fam,KERNEL=finite}).
Unfortunately, the method of using UPB to produce such $\r$ works
only when $M=N=3$. This follows from the following simple lemma.

 \bl
 \label{le:UPBgeneralposition}
If a UPB consists of $(M-1)(N-1)+1$ product vectors in general
position, then $M=N=3$.
 \el
 \bpf
Let $\{\ket{a_i,b_i}:i=1,\ldots,(M-1)(N-1)+1\}$ be a UPB. Since
there are no UPB when $M=2$ or $N=2$, we have $M,N\ge3$. Assume the
$\ket{a_i,b_i}$ are in general position. Let $p$ $[q]$ be the number
of indexes $i$ such that $\ket{a_1}\perp\ket{a_i}$
$[\ket{b_1}\perp\ket{b_i}]$. Our assumption implies that $p\le M-1$
and $q\le N-1$. Since $\ket{a_1,b_1}\perp\ket{a_i,b_i}$ for $i>1$,
we have
 \bea
(M-1)(N-1)\le p+q\le M+N-2.
 \eea
Thus $(M-2)(N-2)\le1$ and so $M=N=3$.
 \epf

\begin{example}
\label{Besk-Fam,KERNEL=finite} $(M=3,N=4)$ {\rm We shall construct
a good real $3\times4$ extreme PPTES $\r$ of birank $(5,5)$ such that
$\cR(\r)$ is a CES and the 10 product vectors belonging to $\ker\r$
are in general position.

We write $\r$ as in Eq. (\ref{MxN-State}) where we set $M=3$, $N=4$,
$R=5$ and define the blocks $C_i$ by
 \bea \label{PrimMatC1,KERNEL=finite}
 C_0= \left[\begin{array}{cccc}
1 & 0 & 0 & 0 \\
0 & 1 & 0 & 0 \\
0 & 0 & 1 & 0 \\
0 & 0 & 0 & 0 \\
0 & 0 & 0 & 0
\end{array}\right],\quad
 C_1= \left[\begin{array}{rrrr}
0 & 1 & 2  & 0 \\
1 & 0 & 0  & 0 \\
2 & 0 & 1  & 0 \\
0 & 0 & 1  & 0 \\
0 & 0 & -1 & 1
\end{array}\right],\quad
 C_2= \left[\begin{array}{rrrr}
1  & 0  & 0  & 0 \\
0  & -1 & 1  & 0 \\
0  & 1  & -1 & 0 \\
-3 & -1 & 1  & 1 \\
0  & 0  & 0  & 1
\end{array}\right].
 \eea
One can verify by direct computation that there are exactly 10
product vectors in $\ker\r$, and that they are in general position.
Moreover, $\cR(\r)$ is a CES and $\r$ is entangled. Since $C_0^\dag
C_1$, $C_0^\dag C_2$ and $C_1^\dag C_2$ are real symmetric matrices,
we have $\r^\G=\r$. Hence $\r$ is PPT and $\rank\r^\G=5$. By Theorem
\ref{thm:MxNrank(N+1)reducible} $\r$ is irreducible, and by Theorem
\ref{thm:3xN-PPTESbirank} it is extreme. \hfill $\square$ }
\end{example}

Let $\ket{a_i,b_i}$, $i=1,\ldots,10$, be the product vectors in this
example. According to Lemma \ref{le:UPBgeneralposition}, there is no
ILO $A\ox B$ such that some seven of the ten product vectors
$A\ox B\ket{a_i,b_i}$ (after normalization) form an UPB.
As $\cR(\r)$ is a CES and the $\ket{a_i,b_i}$ span $\ker\r$,
we can say that $\ker\r$ is spanned
by a general UPB according to the following definition.
A  {\em general UPB} is a set of linearly independent product vectors
$\{\psi\}:=\{\ket{\psi_i}:i=1,\ldots,k\}\subset\cH$ such that
$\{\psi\}^\perp$ is a CES \cite{skowronek11}.

\begin{example} \label{Besk-Fam,KERNEL=finite,M=3,Nge4}
$(N>M=3)$ {\rm We shall construct a family of good real $3\times N$
extreme states $\r$ of birank $(N+1,N+1)$ depending on $N-3$
parameters. By Theorem \ref{thm:LMS2010(ii),ker=finitePRO}, these
states are strongly extreme.

We write $\r$ as in Eq. (\ref{MxN-State}) where we set $M=3$,
$R=N+1$ and define the $R\times N$ blocks $C_0=I_{N-1}\oplus0$ and
 \bea \label{PrimMatC2,KERNEL=finite}
 C_1 = \left[\begin{array}{crrr}
(N-3)(B^2-I_{N-3}) & {\bf e} & {\bf b} & 0 \\
{\bf e}^T          &    0    &    0    & 0 \\
{\bf b}^T          &    0    &    1    & 0 \\
0 & 0 &  1 & 0 \\
0 & 0 & -1 & 1
\end{array}\right], \quad
 C_2 = \left[\begin{array}{crrr}
B-I_{N-3} & 0 & 0 & 0  \\
0 & -1 &  1 & 0 \\
0 &  1 & -1 & 0 \\
{\bf e}^T-{\bf b}^T B & -1 &  1 & 1 \\
0 & 0 & 0 & 1
\end{array}\right],
 \eea
where {\bf e} is the all-one column vector, {\bf b} column vector
with components $b_1,\ldots,b_{N-3}$, and $B$ the diagonal matrix
with diagonal entries $b_1,\ldots,b_{N-3}$. We assume that
the $b_i$ are real, $b_i^2\ne 1$, and that the $b_i^2$ are
pairwise distinct.

We claim that $\cR(\r)$ is a CES. To prove this claim, it suffices
to show that the matrix
 \bea
\left[\begin{array}{c}\x C_0\\ \x C_1\\ \x C_2 \end{array}\right]
 \eea
cannot have rank one for any row vector $\x\in\bC^{N+1}$. We omit
the straightforward and tedious verification.

It follows from this claim that $\r$ is entangled.
Since $C_0^\dag C_1$, $C_0^\dag C_2$ and $C_1^\dag C_2$
are real symmetric matrices, we have $\r^\G=\r$. Hence $\r$ is a
PPTES of birank $(N+1,N+1)$. This state is irreducible by Theorem
\ref{thm:MxNrank(N+1)reducible}, and extreme by Theorem
\ref{thm:3xN-PPTESbirank}.
For $N=4,5,6$ and several choices of the parameters $b_i$, P. Sollid
\cite{sollid12} has verified computationally that the $\d$
product vectors in $\ker\r$ are in general position.
Whether this is true in general, remains an open question.
 \hfill $\square$ }
\end{example}

 \bt \label{thm:PrvaFam}
All $3\times N$ PPT states $\r$ constructed in
Example \ref{Besk-Fam,KERNEL=finite,M=3,Nge4} are good.
 \et
 \bpf Since $\r$ has rank $M+N-2$ $(=N+1)$, it suffices to prove
that $X_\r$ is a finite set.

Assume that $|X_\r|=\infty$. By Theorem \ref{thm:delta}, there
is a 2-dimensional subspace $V\ox W\se\ker\r$. Since $\r$ is
irreducible, Proposition \ref{prop:MxNrank(N+1),commonkernel,PPT}
implies that we must have $\dim V=1$ and $\dim W=2$. Let
$\x_0\ket{0}+\x_1\ket{1}+\x_2\ket{2}\in V$ be a nonzero vector.
Then the matrix $\Xi=\x_0C_0+\x_1C_1+\x_2C_2$ has rank less than
$N-1$. Obviously, at most one $\x_i$ is 0. Suppose that $\x_1=0$.
Then the $(N-3)\times3$ submatrix in the upper right corner of
$\Xi$ is 0, the $4\times3$ submatrix below it has rank 3, and the
$(N-3)\times(N-3)$ submatrix in the upper left corner of $\Xi$ has
rank $\ge N-4$, and we have a contradiction. Hence, $\x_1\ne0$
and we can assume that $\x_1=1$. Thus we have
 \bea
 \Xi = \left[\begin{array}{cccc}
\x_0 I_{N-3}+(N-3)(B^2-I_{N-3})+\x_2(B-I_{N-3})
                            & {\bf e}   & {\bf b}     & 0     \\
{\bf e}^T                   & \x_0-\x_2 &   \x_2      & 0     \\
{\bf b}^T                   & \x_2      & 1+\x_0-\x_2 & 0     \\
\x_2({\bf e}^T-{\bf b}^T B) & -\x_2     &   1+\x_2    & \x_2  \\
0                           &    0      &   -1        & 1+\x_2
\end{array}\right].
 \eea

Suppose that $\x_2=0$. Then the $2\times(N-2)$ submatrix in the
lower left corner of $\Xi$ is 0 and the $2\times2$ submatrix in the
lower right corner has rank 2. Since the $b_i^2$ are pairwise
distinct, the $(N-3)\times(N-2)$ submatrix in the upper left corner
of $\Xi$ has rank $N-3$, and we have a contradiction. Hence,
$\x_2\ne0$.

Suppose that $\x_2=-1$. Then $\rank \Xi=2+r$, where $r$ is the rank of
the $(N-1)\times(N-2)$ submatrix in the upper left corner of $\Xi$. It
is easy to show that $r\ge N-3$, and so we have again a
contradiction. Hence, $\x_2\ne-1$.

At most two of the first $N-3$ diagonal entries
$d_i:=(N-3)(b_i^2-1)+\x_0+(b_i-1)\x_2$, $i=1,\ldots,N-3$, of $\Xi$ may
be 0. Suppose that exactly one of the $d_i$ is 0, say $d_1=0$. Then,
by using the fact that $\x_2\ne-1$, we see that the
$(N-1)\times(N-1)$ submatrix of $\Xi$ obtained by removing the rows
$N-1$ and $N$ and column $N-1$ is invertible, and we have a
contradiction. Suppose that exactly two of the $d_i$ are 0, say
$d_1=d_2=0$. Then the $(N-1)\times(N-1)$ submatrix in the upper left
corner of $\Xi$ is invertible, and we have a contradiction. We
conclude that all $d_i$ are nonzero.

For $i\in\{N-2,N-1,N\}$, let $\Xi[i]$ denote the square submatrix of
$\Xi$ of order $N-2$ obtained by deleting the last two columns and
keeping only the $i$th row from the last four rows. Since
$\Xi[N-2]\oplus[1+\xi_2]$ is a $(N-1)\times(N-1)$ submatrix of $\Xi$ and
$\xi_2\ne-1$, our assumption implies that $\det \Xi[N-2]=0$.
Similarly, one can show that $\det \Xi[N-1]=\det \Xi[N]=0$. Hence we
have
 \bea
 \label{ea:3xNrank(N+1),finite1}
\x_0-\x_2=\sum_i \frac{1}{d_i},\quad
 \x_2=\sum_i\frac{b_i}{d_i},\quad 1=\sum_i\frac{b_i^2-1}{d_i}.
 \eea
By multiplying these equations by $\x_0-\x_2$, $\x_2$, $N-3$
respectively and adding them up, we obtain that
$(\x_0-\x_2)^2+\x_2^2=0$. Hence $\x_0$ or $\x_2$ is not real. Let us
denote by $\x'_0$, $\x'_2$, $d'_i$ the imaginary parts of
 $\x_0$, $\x_2$, $d_i$, respectively.
From the definition of the $d_i$ we have $d'_i=\x'_0+(b_i-1)\x'_2$.
From Eq. (\ref{ea:3xNrank(N+1),finite1}) we have
 \bea
\x'_0-\x'_2=-\sum_i \frac{d'_i}{|d_i|^2},\quad
 \x'_2=-\sum_i \frac{b_id'_i}{|d_i|^2}.
 \eea
By multiplying the first of these equations by $\x'_0-\x'_2$ and the
second by $\x'_2$ and adding them, we obtain that
 \bea
(\x'_0-\x'_2)^2+(\x'_2)^2=-\sum_i \frac{(d'_i)^2}{|d_i|^2}.
 \eea
As the left hand side is positive, we have a contradiction. Hence,
our assumption is false and so $\r$ is good.
 \epf

\subsection{Bad case: infinitely many product vectors in the kernel}

The examples in this subsection will be all bad, i.e. the kernel of
the state will contain infinitely many product vectors. The Examples
\ref{Besk-Fam}, \ref{Besk-Niz} and \ref{eg:Mge3,Nge4,extreme} cover
all possible local ranks $(M,N)$, except $M=N=3$ which is an
exception (see \cite[Theorem 22]{cd11JMP}). The first two examples
are easily shown to be extreme. Moreover, we prove that all states
in Example \ref{eg:Mge3,Nge4,extreme} are extreme (see Theorem
\ref{thm:MxNrank(M+N-2)}) and thereby we confirm part (i) of
Conjecture \ref{conj:LMS2010highdimension}.

Since UPBs are used extensively in quantum information, we first
consider the PPTES $\r$ associated to an arbitrary UPB of the family
{\bf GenTiles2}.
Suppose that $N\ge M\ge3$ and $N>3$. Then by \cite[Theorem 6]{DiV03}
the following $MN-2M+1$ o.n. product vectors form a UPB:
 \begin{eqnarray}
 && \ket{S_j}:=\frac{1}{\sqrt2}(\ket{j}-\ket{j+1 \mod M})
\ox\ket{j}, \\
 &&  0\le j\le M-1; \\
 &&  \ket{L_{jk}} := \ket{j} \ox  \frac{1}{\sqrt {N-2}}
 \bigg( \sum^{M-3}_{i=0} \o^{ik} \ket{i+j+1 \mod M} +
 \sum^{N-3}_{i=M-2} \o^{ik} \ket{i+2} \bigg), \\
 && \o:=e^\frac{2\pi i}{N-2},\quad 0\le j\le M-1,
 \quad 1\le k\le N-3; \\
 &&  \ket{F} := \frac{1}{\sqrt{NM}}
 \sum^{M-1}_{i=0} \sum^{N-1}_{j=0} \ket{i}\ox\ket{j}.
 \end{eqnarray}

 \bl \label{le:UPB}
Let $\r$ be the PPTES of rank $2M-1$ associated with this UPB, i.e.,
 \bea \label{ro:GenTiles2}
 \r = I_{MN}-\sum^{M-1}_{j=0}\proj{S_j}-\sum^{M-1}_{j=0}
 \sum^{N-3}_{k=1}\proj{L_{jk}}-\proj{F}.
 \eea
Then $\rank\r_A=M$ and, if $N>M$, $\rank\r_B=M+1$.
 \el

 \bpf
By a direct tedious computation, which we omit, we find that
$\r_A=Z/2M$, where $Z$ is the circulant matrix with the first row
$[4M-2,M-2,-2,-2,\ldots,-2,M-2]$. Hence $\det Z=\prod_\z f(\z)$,
where the product is taken over all $M$th roots of unity, $\z$, and
$f(t)$ is the polynomial
\begin{eqnarray} \notag
f(t) &=& 4M-2+(M-2)t-2t^2-2t^3-\cdots-2t^{M-2}+(M-2)t^{M-1} \\
&=& M(4+t+t^{M-1})-2(1+t+t^2+\cdots+t^{M-1}).
\end{eqnarray}
Since $f(1)=4M$ and $f(\z)=M(4+\z+\z^{-1})$ when $\z^M=1$ but
$\z\ne1$, all of these numbers are nonzero, and so $\rank\r_A=M$.

Now assume that $N>M$. By another straightforward tedious
computation, we find that
 \bea
 \r_B=\frac{1}{N(N-2)} \left[ \begin{array}{cc}
 U & X \\ X^T & Y \end{array} \right]=\frac{1}{N(N-2)}W,
 \eea
where $U$ is a circulant matrix of order $M$ with first row
 \bea
 [(M+N-5)N+2,(M-4)N+2,(M-5)N+2,\ldots,(M-5)N+2,(M-4)N+2],
 \eea
and $X$ and $Y$ are matrices all of whose entries are equal to
$(M-3)N+2$ and $(M-1)N+2$, respectively. Since the last $N-M$ rows
of $W$ are all equal to each other, it is clear that $\rank\r_B\le
M+1$.

It remains to show that the matrix $V$ of order $M+1$, obtained
by deleting the last $N-M-1$ rows and columns of $W$, is
nonsingular. By subtracting $\l:=((M-3)N+2)/((M-1)N+2)$ times
the last column of $V$ from all other columns, the problem
reduces to proving that the matrix $U':=(M-1+2/N)(U-\l J)$ is
nonsingular, where $J$ is all-ones matrix. The matrix $U'$ is
also circulant with first row
 \bea
 [N((M-1)N-2),(M-5)N+2,-4N,-4N,\ldots,-4N,(M-5)N+2].
 \eea
We can now prove that $U'$ is nonsingular by using the same argument
as for $Z$. \epf

The state $\r$ defined by Eq. (\ref{ro:GenTiles2}) is $\G$-invariant
and, for $N>4$, contains infinitely many product vectors in its
kernel. Both assertions are immediate from the definition of the
product vectors $S_j$, $L_{jk}$ and $F$. Apparently all these states
$\r$ are extreme; we have verifed it in several cases by using the
Extremality Criterion (see Proposition \ref{prop:ExtCrit}). It
follows easily from the above proof that
 \bea
 \det \r_A=2\prod_{k=1}^{M-1}\left( 2+\cos\frac{2\pi k}{M} \right).
 \eea

\begin{example} \label{Besk-Fam}
$(M=3,N=4)$ {\rm We shall construct a real 7-parameter family
of $3\times4$ extreme PPTES $\r$ of birank $(5,5)$ such that
$\cR(\r)$ is a CES and $\ker\r$ contains infinitely many product
vectors.

The states $\r$ in this family are given by Eq. (\ref{MxN-State})
with $M=3$, $N=4$ and $R=5$. The blocks $C_i$ are given by
 \bea \label{PrimMatC1}
 C_0= \left[\begin{array}{cccc}
1 & 0 & 0 & 0 \\
0 & 1 & 0 & 0 \\
0 & 0 & 0 & 0 \\
0 & 0 & 0 & 0 \\
0 & 0 & 0 & 0
\end{array}\right],\quad
 C_1= \left[\begin{array}{cccc}
0 & 0 & 0 & 0 \\
0 & -be/a & 0 & 0 \\
0 & 0 & 1 & 0 \\
0 & 0 & 0 & 1 \\
0 & b & 0 & 0
\end{array}\right],\quad
 C_2= \left[\begin{array}{cccc}
0 & a & 0 & 0 \\
a & f & 0 & 0 \\
0 & b & 0 & 0 \\
0 & bd & 0 & c \\
e & g & 1 & d
\end{array}\right],
\eea where $a,b,c,d,e,f,g$ are real parameters which are all nonzero
except possibly $f$ and $g$.

Since $\ket{0,2},\ket{0,3}\in\ker\r$, it is obvious that $\r$ is bad.
If we identify $\cH_A\ox\cH_B$ with the space of $3\times4$ complex
matrices, then the five rows of $C=[C_0~ C_1~ C_2]$ are represented
by the matrices
 \bea
\left[\begin{array}{cccc}
1 & 0 & 0 & 0 \\
0 & 0 & 0 & 0 \\
0 & a & 0 & 0
\end{array}\right],
\left[\begin{array}{cccc}
0 & 1 & 0 & 0 \\
0 & -be/a & 0 & 0 \\
a & f & 0 & 0
\end{array}\right],
\left[\begin{array}{cccc}
0 & 0 & 0 & 0 \\
0 & 0 & 1 & 0 \\
0 & b & 0 & 0
\end{array}\right],
\left[\begin{array}{cccc}
0 & 0 & 0 & 0 \\
0 & 0 & 0 & 1 \\
0 & bd & 0 & c
\end{array}\right],
\left[\begin{array}{cccc}
0 & 0 & 0 & 0 \\
0 & b & 0 & 0 \\
e & g & 1 & d
\end{array}\right].
 \eea
It is obvious that these matrices are linearly independent, and so
$\rank\r=5$. It is easy to verify that the space spanned by these
five matrices contains no matrix of rank one. Consequently,
$\cR(\r)$ is a CES and $\r$ is entangled. Since $C_0^\dag C_1$,
$C_0^\dag C_2$ and $C_1^\dag C_2$ are real symmetric matrices, we
have $\r^\G=\r$. Hence $\r$ is PPT and $\rank\r^\G=5$. By Theorem
\ref{thm:MxNrank(N+1)reducible} $\r$ is irreducible, and by Theorem
\ref{thm:3xN-PPTESbirank} it is extreme. \hfill $\square$ }
\end{example}

We now specialize the values of the parameters in the above example
and extend this particular case to obtain $3\times N$ PPTES
$\r^{(N)}$ for all $N\ge4$. Each state $\r^{(N)}$ is extreme,
$\G$-invariant, has rank $N+1$, its range is a CES, and its kernel
contains infinitely many product vectors.

\begin{example} \label{Besk-Niz}
$(N>M=3)$ {\rm Let us denote by $C^{(4)}_i$, $i=0,1,2$, the
matrices (\ref{PrimMatC1}) where we set $a=b=c=d=e=1$ and
$f=g=0$.
For $N>4$ we define the $(N+1)\times N$ matrices $C^{(N)}_i$,
$i=0,1,2$, as follows: $C^{(N)}_0=I_{N-4}\oplus C^{(4)}_0$ and
 \bea \label{3xN-MatC}
 C^{(N)}_1= \left[\begin{array}{ccccccccc}
0 & \ddots & 0 & 0 & 0 & 0 & 0 & 0 & 0 \\
\ddots & 0 & 1 & 0 & 0 & 0 & 0 & 0 & 0 \\
0 & 1 & 0 & 1 & 0 & 0 & 0 & 0 & 0 \\
0 & 0 & 1 & 0 & 1 & 0 & 0 & 0 & 0 \\
0 & 0 & 0 & 1 & 0 & 1 & 0 & 0 & 0 \\
0 & 0 & 0 & 0 & 1 & 0 & 0 & 0 & 0 \\
0 & 0 & 0 & 0 & 0 & 0 & -1 & 0 & 0 \\
0 & 0 & 0 & 0 & 0 & 0 & 0 & 1 & 0 \\
0 & 0 & 0 & 0 & 0 & 0 & 0 & 0 & 1 \\
0 & 0 & 0 & 0 & 0 & 0 & 1 & 0 & 0
\end{array}\right], \quad
 C^{(N)}_2= \left[\begin{array}{ccccccccc}
0 & \ddots & 0 & 0 & 0 & 0 & 0 & 0 & 0 \\
\ddots & 0 & 1 & 0 & 0 & 0 & 0 & 0 & 0 \\
0 & 1 & 0 & 1 & 0 & 0 & 0 & 0 & 0 \\
0 & 0 & 1 & 0 & 1 & 0 & 0 & 0 & 0 \\
0 & 0 & 0 & 1 & 0 & 1 & 0 & 0 & 0 \\
0 & 0 & 0 & 0 & 1 & 0 & 1 & 0 & 0 \\
0 & 0 & 0 & 0 & 0 & 1 & 0 & 0 & 0 \\
0 & 0 & 0 & 0 & 0 & 0 & 1 & 0 & 0 \\
0 & 0 & 0 & 0 & 0 & 0 & 1 & 0 & 1 \\
0 & 0 & 0 & 0 & 1 & 1 & 0 & 1 & 1
\end{array}\right].
 \eea
Let
$\r^{(N)}=\sum_{i,j}\ket{i}\bra{j}\ox C^{(N)\dag}_i C^{(N)}_j$,
$N\ge4$.
It is not hard to verify that $\r^{(N)}$ is a $3\times N$ state.
Indeed, we have
\bea \r_A^{(N)}=\left[\begin{array}{ccc} N-2
& -1 & 0 \\ -1 & 2N-4 & 2N-7 \\ 0 & 2N-7 & 2N+1
\end{array}\right]>0.
\eea
Since $C^{(N)\dag}_1 C^{(N)}_1=S\oplus\diag(2,1,1)$ with $S\ge0$,
and  $C^{(N)\dag}_0 C^{(N)}_0=I_{N-3}\oplus\diag(0,0,0)$,
we also have $\r_B^{(N)}>0$.

The matrices $C^{(N)\dag}_i C^{(N)}_j$, $i<j$, are real and
symmetric, and so $\r^{(N)}$ is $\G$-invariant. In particular,
$\r^{(N)}$ is a PPT state. Since the last two columns of
$C^{(N)}_0$ are 0, we have
$\ket{0}\ox(\xi\ket{N-2}+\eta\ket{N-1})\in\ker\r^{(N)}$ for all
$\xi,\eta\in\bC$.

We claim that $\r^{(N)}$ has rank $N+1$ and that its range is a CES.
Let $C^{(N)}=[C^{(N)}_0~C^{(N)}_1~C^{(N)}_2]$. Each column of the
$3N\times(N+1)$ matrix $C^{(N)\dag}$ represents a vector in
$\cH$.
These vectors span the range of $\r^{(N)}$. We can represent
these vectors by $3\times N$ matrices $W_1,\ldots,W_{N+1}$.
It is easy to see that these matrices are linearly independent,
and so $\r^{(N)}$ has rank $N+1$. An arbitrary vector in
$\cR(\r^{(N)})$ is represented by a matrix $\sum_i \xi_i W_i$,
where the $\xi_i$ are complex scalars. It is now easy to verify
that this matrix cannot have rank 1, i.e., $\cR(\r^{(N)})$ is a CES.

As in the previous example, it follows that $\r^{(N)}$ is extreme.
\hfill $\square$ }
\end{example}

We now construct a new family of examples of PPTES which extends
the one-parameter family $\r^{(N)}$, $N\ge4$, of Example
\ref{Besk-Niz}. This new family depends on two discrete parameters
$M$ and $N$, and $M-3$ real parameters $c_i$, $i=3,4,\ldots,M-1$.
It consists of $M\times N$ PPTES $\r^{(M,N)}$ of rank $M+N-2$.
Note that $\r^{(3,N)}=\r^{(N)}$. We will prove that each
state $\r^{(M,N)}$ is $\G$-invariant, its range is a CES, and its
kernel contains infinitely many product vectors. By using the
Extremality Criterion (Proposition \ref{prop:ExtCrit}), we have
verified that they are extreme for $M+N\le27$.
We shall prove in Theorem \ref{thm:MxNrank(M+N-2)} below that
all of these states are extreme.

 \begin{example} \label{eg:Mge3,Nge4,extreme}
$(M\ge3,N\ge4)$ {\rm We define the $(M+N-2)\times N$ matrices
 \bea
\label{MxN-MatC} C^{(M,N)}_i &=& \left[\begin{array}{c} C_i^{(N)} \\
Q_i
\end{array}\right], \quad i=0,1,2; \\
\label{MxN-MatC,2} C^{(M,N)}_i &=& \left[\begin{array}{c} P \\ Q_i
\end{array}\right], \quad i=3,4,\ldots,M-1;
 \eea
where the $C_i^{(N)}$, $i=0,1,2$, are given by Eqs.
(\ref{3xN-MatC}); the $(1,2)$ entry of $P$ is 1 and all other are 0;
the first column of $Q_0$ has all entries equal to 1 and all other
columns are 0; $Q_1=Q_2=0$ and for $i>2$ each $Q_i$ has exactly two
nonzero entries, namely $(i-2,1)$th entry is $c_i$ and $(i-2,2)$th
entry is $-1$. The numbers $c_i$, $i=3,4,\ldots,M-1$ are required to
be real, nonzero and distinct.

Let $\r^{(M,N)}=C^\dag C$ where
 \bea \label{MatCMxN}
C:=[~C^{(M,N)}_0~C^{(M,N)}_1~\cdots~C^{(M,N)}_{M-1}~].
 \eea

It is not hard to verify that $\r^{(M,N)}$ is an $M\times N$ state
of rank $M+N-2$. The fact that $\rank\r^{(M,N)}_B=N$ follows from
Example \ref{Besk-Niz} where we have shown that $\rank\r^{(N)}_B=N$.
To prove that $\rank\r^{(M,N)}_A=M$, we first compute the reduced
density matrix \bea \r_A^{(M,N)}=\left[\begin{array}{ccc|cccc}
M+N-5 & -1 & 0 & c_3 & c_4 & \cdots & c_{M-1} \\
-1 & 2N-4 & 2N-7 & 1 & 1 &  & 1 \\
0 & 2N-7 & 2N+1 & 1 & 1 &  & 1 \\
\hline
c_3 & 1 & 1 & 2+c_3^2 & 1 &  & 1 \\
c_4 & 1 & 1 & 1 & 2+c_4^2 &  & 1 \\
\vdots &&&&&& \\
c_{M-1} & 1 & 1 & 1 & 1 &  & 2+c_{M-1}^2
\end{array}\right].
\eea Since
$\left[\begin{array}{cc}1&c_i\\c_i&c_i^2\end{array}\right]\ge0$ for
each $i$, it suffices to show that \bea \label{PozDefJ}
\left[\begin{array}{cc}
\r^{(N)}_A & E^T \\
E & I_{M-3}+J_{M-3}
\end{array}\right]>0,
\eea where $\r^{(N)}_A$ is as in Example \ref{Besk-Niz}, $J_{M-3}$
is all-ones matrix, and so is $E$ except that its first column is 0.
By using \cite[Proposition 8.2.3]{bernstein05book} and the fact that
$(I_{M-3}+J_{M-3})^{-1}=I_{M-3}-J_{M-3}/(M-2)$, one deduces that the
inequality (\ref{PozDefJ}) is equivalent to \bea
\left[\begin{array}{ccc}
N-2 & -1 & 0 \\
-1 & 2N-5+\frac{1}{M-2} & 2N-8+\frac{1}{M-2} \\
0 & 2N-8+\frac{1}{M-2} & 2N+\frac{1}{M-2}
\end{array}\right]>0.
\eea It suffices to verify the latter inequality for $M=+\infty$
only, which is straightforward. Finally, to prove that
$\rank\r^{(M,N)}=M+N-2$, we have to show that $C$ has full rank. It
follows from Example \ref{Besk-Niz} that the first $N+1$ rows of $C$
are linearly independent. Then by inspecting the first columns of
$C^{(M,N)}_i$, $3\le i\le M-1$, we see that $C$ indeed has full
rank.

One can easily verify that the matrices $C^{(M,N)\dag}_i
C^{(M,N)}_j$, $i<j$, are symmetric. Since they are also real, it
follows that $\r^{(M,N)}$ is $\G$-invariant, and so $\r^{(M,N)}$
is a PPT state.

Next we claim that the range of $\r^{(M,N)}$ is a CES. Each
column of $C^\dag$ represents a vector in $\cH$. These $M+N-2$
vectors span the range of $\r^{(M,N)}$. We can represent these
vectors by $M\times N$ matrices $W_1,\ldots,W_{M+N-2}$, respectively. An
arbitrary vector in $\cR(\r^{(M,N)})$ is represented by a linear
combination $\Xi=\sum_i \xi_i W_i$, where the $\xi_i$ are complex
scalars. Now it suffices to verify that $\Xi$ cannot have rank 1.
We briefly indicate the main steps. It follows from Example
\ref{Besk-Niz} that the claim holds if $\x_i=0$ for $i>N+1$.
Otherwise we must have $\x_i=0$ for $4\le i\le N-2$. Next one can
deduce that $\x_{N-1}=\x_{N}=\x_{N+1}=0$, and then that also
$\x_{1}=\x_{2}=\x_{3}=0$. Thus only the variables
$\x_{N+2},\ldots,\x_{M+N-2}$ may be nonzero. However at least
$M-4$ of them, say $\x_{N+2},\ldots,\x_{M+N-3}$ have to vanish
because the $c_i$ are distinct. As $\rank W_i>1$ for each $i$,
the last coefficient $\x_{M+N-2}$ must also vanish.

Since the last two columns of $C^{(M,N)}_0$ are 0, we have
$\ket{0}\ox(\xi\ket{N-2}+\eta\ket{N-1})\in\ker\r^{(M,N)}$ for all
$\xi,\eta\in\bC$. Thus $\r^{(M,N)}$ is bad.
 \hfill $\square$ }
 \end{example}

 \bt \label{thm:MxNrank(M+N-2)}
For $M,N>2$ we have $\cE_{M+N-2}^{M,N}\ne\emptyset$, i.e., part
(i) of Conjecture \ref{conj:LMS2010highdimension} is valid.
 \et
 \bpf
This is well known for $M=N=3$. It suffices to prove that the
states $\r=\r^{(M,N)}$ defined in
Example \ref{eg:Mge3,Nge4,extreme} are extreme when $M\le N$.
For $M=3$ this was shown in Example \ref{Besk-Niz}.
Hence, we may assume that $M\ge4$.

For convenience, we set $R=M+N-2$ and we switch the two blocks in
Eqs. (\ref{MxN-MatC}) and (\ref{MxN-MatC,2}). Thus we define the
$R\times N$ blocks
 \bea
\label{MxN-MatrCi} C_i&=&\left[\begin{array}{c} Q_i\\C_i^{(N)}
\end{array}\right], \quad i=0,1,2; \\
\label{MxN-MatrCii} C_i&=&\left[\begin{array}{c} Q_i \\ P
\end{array}\right], \quad i=3,4,\ldots,M-1.
 \eea
The equality $\r=C^\dag C$ remains valid, with
$C:=[C_0~C_1~\cdots~C_{M-1}]$.

Let $\s$ be a PPT state such that $\cR(\s)=\cR(\r)$ and
$\cR(\s^\G)=\cR(\r^\G)$. Recall that $\r^\G=\r$. By Proposition
\ref{prop:ExtCrit}, it suffices to show that $\s$ must be a scalar
multiple of $\r$. Since $\r$ and $\s$ have the same range, there
exists an invertible matrix $A$ such that
 \bea
\s=(A C)^\dg A C=C^\dag HC,
 \eea
where $H=[h_{ij}]:=A^\dag A>0$. Let us partition $\s$ into $M^2$
blocks $\s=[\s_{rs}]_{r,s=0}^{M-1}$ where $\s_{rs}:=C_r^\dag HC_s$.
We use $r$ $[s]$ as the blockwise row [column] index. Then
$\s^\G=[\s_{sr}]_{r,s=0}^{M-1}$ is the blockwise transpose of $\s$.

Let us partition the index set $\{1,2,\ldots,R\}$ into three
subsets $J_1,J_2,J_3$ defined by inequalities $i\le M-2$,
$M-2<j<R-2$, $k\ge R-2$, respectively.

The last two columns of $C_0$ are 0. Hence, the last two diagonal
entries of $\s_{00}=C_0^\dag HC_0$ are 0. Since $\s^\G\ge0$, the
last two rows of each block $\s_{s0}=C_s^\dag HC_0$ must be 0. By
taking $s=1,2$ we see that $h_{ij}=0$ for $i\in J_2$ and $j\in J_3$.

Assume that the index $s>2$. Then all but the first two columns of
$C_s$ are 0. Hence, all diagonal entries but the first two of
$\s_{ss}$ are 0. Consequently, the last $N-2$ rows of the blocks
$\s_{rs}$, $r=1,2$, must be 0. Equating the last two rows of these
blocks to 0, we see that $h_{ij}=0$ for $i\in J_1$ and $j\in J_3$. If
$N>4$ then the equations provided by the middle $N-4$ rows of the
same blocks imply that $h_{ij}=0$ for $i\in J_1$ and $j\in J_2$. The
same conclusion is valid in the case $N=4$ because in that case we
have $J_2=\{M-1\}$ and the first column of $C_1$ is 0, so the
$(N+1)$th row of $\s^\G$ must be 0.

Thus $H=H_1\oplus H_2\oplus H_3$, with square matrices $H_1,H_2,H_3$
of order $M-2$, $N-3$ and 3, respectively.

Let $V_k$, $1\le k\le R$, be the $M\times N$ matrix whose $i$th
row is the $k$th row of $C_{i-1}$, $i=1,2,\ldots,M$. These
matrices represent vectors in $\cH$ and as such they form a basis
of $\cR(\r)$. An arbitrary vector in $\cR(\r)$ is represented by
the linear combination $\Xi:=\sum_k \xi_k V_k$, where $\xi_k$ are
complex scalars. Explicitly,

 \bea
\Xi=\left[
 \begin{array}{ccccc}
 \xi_1+\cdots+\xi_{M-2}& \xi_{M-1} & \xi_M & \quad\cdots\quad &
 \xi_{R-6} \\
 \xi_{M-1} & \xi_{M-2}+\xi_M & \xi_{M-1}+\xi_{M+1} & &
 \xi_{R-7}+\xi_{R-5} \\
 \xi_{M-1} & \xi_{M-2}+\xi_M & \xi_{M-1}+\xi_{M+1} & &
 \xi_{R-7}+\xi_{R-5} \\
c_3 \xi_1 & \xi_{M-2}-\xi_1 & 0 & & 0 \\
 \vdots & & & & \\
 c_{M-1} \xi_{M-3} & \xi_{M-2}-\xi_{M-3} & 0 & & 0 \\
 \end{array}
\right. \qquad\qquad\qquad\qquad\qquad\qquad
\notag \\
\label{DugMat} \qquad \left.
 \begin{array}{ccccc}
 \xi_{R-5} & \xi_{R-4} & \xi_{R-3} & 0 & 0 \\
 \xi_{R-6}+\xi_{R-4} & \xi_{R-5} & \xi_R-\xi_{R-3} & \xi_{R-2} &
 \xi_{R-1} \\
 \xi_{R-6}+\xi_{R-4}+\xi_R & \xi_{R-5}+\xi_{R-3}+\xi_R
 & \xi_{R-4}+\xi_{R-2}+\xi_{R-1} & \xi_R & \xi_{R-1}+\xi_R \\
 0 & 0 & 0 & 0 & 0 \\
 \vdots & & & & \\
 0 & 0 & 0 & 0 & 0 \\
 \end{array}
\right].
 \eea

The range of the matrix $\s^\G$ is spanned by its columns, which we
represent by $M\times N$ matrices. Let $S_k$ be the $M\times N$
matrix representing the $k$th column of $\s^\G$.

We have
 \bea
S_1:=\left[
  \begin{array}{cccccccc}
 \sum H_1 & 0 & 0 & \quad\cdots\quad & 0 & 0 & 0 & 0  \\
 0 & h_{M-1,M-1} & h_{M,M-1} & & h_{R-4,M-1} &  h_{R-3,M-1} & 0 & 0 \\
 0 & h_{M-1,M-1} & h_{M,M-1} & & h_{R-4,M-1} &  h_{R-3,M-1} & 0 & 0 \\
 c_3 H_1(1) & 0 & 0 & & 0 & 0 & 0 & 0 \\
 \vdots & & & & & & & \\
 c_{M-1} H_1(M-3) & 0 & 0 & & 0 & 0 & 0 & 0 \\
  \end{array}
\right],
 \eea
where $H_1(k)$ is the sum of the entries in the $k$th column of
$H_1$ and $\sum H_1$ is the sum of all entries of $H_1$. Since
$\cR(\s^\G)=\cR(\r^\G)$, this matrix must have the form
(\ref{DugMat}). From the equation $S_1=\Xi$ we obtain that $\xi_i=0$
for $i\ge M-1$, $\xi_1=\xi_2=\cdots=\xi_{M-2}$ as well as that all
off-diagonal entries of the first row of $H_2$ are 0, and that
$h_{M-1,M-1}=H_1(1)=H_1(2)=\cdots=H_1(M-2)$.

Next we have
 \bea
S_2:=\left[
  \begin{array}{ccccccccc}
 0 & h_{M-1,M-1} & 0 & 0 & \quad\cdots\quad & 0 & 0 & 0 & 0 \\
 0 & 0 & h_{M,M} & h_{M+1,M} & & h_{R-5,M} & h_{R-3,M} & 0 & 0 \\
 0 & 0 & h_{M,M} & h_{M+1,M} & & h_{R-5,M} & h_{R-3,M} & 0 & 0 \\
 0 & 0 & 0 & & 0 & 0 & 0 & 0 & 0 \\
 \vdots & & & & & & & & \\
 0 & 0 & 0 & & 0 & 0 & 0 & 0 & 0 \\
  \end{array}
\right].
 \eea
From the equation $S_2=\Xi$ we obtain that $\xi_i=0$ for $i\ne M-1$,
as well as that all off-diagonal entries of the second row of $H_2$
are 0, and that $h_{M,M}=h_{M-1,M-1}$ (if $N>4$). Similarly, by
using the fact that each of the matrices $S_3,\ldots,S_{N-3}$ must
have the form (\ref{DugMat}), we conclude that $H_2=h_{M-1,M-1}
I_{N-3}$.

From the equation $S_{2N-3}=\Xi$ we obtain that
$h_{R-2,R}=h_{R-1,R}=0$ and $h_{R,R}=h_{M-1,M-1}$. From $S_{2N-1}=\Xi$
we obtain that $h_{R-2,R-1}=0$ and $h_{R-2,R-2}=h_{R,R}$, and from
$S_{2N}=\Xi$ that $h_{R-1,R-1}=h_{R,R}$. Hence $H_2=h_{R,R} I_{N-3}$
and $H_3=h_{R,R} I_3$.

From the equation $S_{2N+1}=\Xi$ we first deduce that $\xi_i=0$ for
all $i\ne M-1,R$, and then that all off-diagonal entries of the last
row of $H_1$  are 0. We also obtain the equations
$\xi_{M-1}=h_{M-2,M-2}$, $\xi_R=h_{M-1,M-1}$, and
$\xi_{M-1}+\xi_R=2h_{M-1,M-1}$ which imply that
$h_{M-2,M-2}=h_{M-1,M-1}$. From the matrix equation $S_{3N+1}=\Xi$ we
first deduce that $\xi_i=0$ for $i\ge M-2$, and then that all
off-diagonal entries of the first row of $H_1$  are 0, and finally
that $h_{1,1}=h_{M-2,M-2}$. From the matrix equation $S_{4N+1}=\Xi$ we
first deduce that $\xi_i=0$ for $i\ge M-2$, and then that all
off-diagonal entries of the second row of $H_1$  are 0, and finally
that $h_{2,2}=h_{M-2,M-2}$. Similarly, by using the equations
$S_{rN+1}=\Xi$, with $r=5,6,\ldots,M-1$, we can deduce that
$H_1=h_{1,1} I_{M-2}$. Since $h_{1,1}=h_{M-1,M-1}$, we have
$H=h_{1,1} I_R$. Thus $\s=h_{1,1}\r$, which completes the proof.
 \epf

\section{\label{sec:application} Some open problems}

The sum of two entangled extreme states is not necessarily an
edge state. We shall construct an example.
Let $\r_1$ be any state belonging to the family
\cite[Eq. 108]{cd11JMP} of $3\times3$ PPTES of rank four depending
on four real parameters. We set $\r_2=I_3\ox P\r_1 I_3\ox P^\dg$,
where $P$ is the cyclic permutation matrix with first row
$[0,0,1]$.
By Theorem \ref{thm:3x3PPTstates}, both $\r_1$ and $\r_2$ are
extreme. One can easily verify that the PPT state $\r=\r_1+\r_2$ is
a $3\times3$ state of birank $(8,8)$. It follows from \cite[Theorem
2.3,(ii)]{kkl11} that $\r$ is not an edge state.

\textbf{Problem 1}. {\em Can the sum of two entangled
extreme states be separable?}

Every separable state is a sum of pure product states, but such
decomposition is not unique in general. (We assume that the
summands are pairwise non-parallel.)
We point out that the good $M\times N$ separable states $\s$ of rank $r\le M+N-2$ have this uniqueness property. Indeed, it
follows from Theorem \ref{thm:goodseparablestate} (ii) that
$\s=\sum^r_{i=1}\proj{\psi_i}$, where the $\ket{\psi_i}$ are
product vectors in general position.
By Lemma \ref{le:GenPos,L<=M+N-2} there are no other product
vectors in $\cR(\s)$. So the above decomposition of $\s$ is
unique. Every PPT state is a sum of extreme states.

\textbf{Problem 2}. {\em Which PPTES have unique decomposition
as a sum of extreme states?}

Since most quantum-information tasks require pure states, the
entanglement distillation (i.e., the task of producing pure
entangled states) is a central problem in quantum information
\cite{bds96}. Mathematically, an entangled state $\r$ is
$n$-distillable under LOCC if there exists a pure state $\ket{\ps}$
of Schmidt rank two such that $\bra{\ps}(\r^{\ox n})^\G\ket{\ps}<0$
\cite{dss00}. A state $\r$ is distillable if it is $n$-distillable
for some positive integer $n$. Otherwise we say that $\r$ is
non-distillable.

It follows easily from this definition that no PPTES is distillable
\cite{hhh98}. It is also believed that entanglement distillation may
fail for some NPT states \cite{dss00}. Nevertheless, the PPTES of
full rank can be used to activate the distillability of any NPT
state \cite{hhh99,vw02}. This means that, for any NPT state
$\r_{A_1B_1}$, there exists a PPTES $\r_{A_2B_2}$ of full rank such
that the bipartite state
$\r_{A_1A_2:B_1B_2}:=\r_{A_1B_1}\ox\r_{A_2B_2}$ is 1-distillable.
Hence, there exists a pure state $\ket{\ps}$ of Schmidt rank two
such that
$\bra{\ps}(\r_{A_1B_1}\ox\r_{A_2B_2})^{\G_{A_1A_2}}\ket{\ps}<0$. We
can write $\r_{A_2B_2}=\sum_{i=1}^k\r_i$, where $\r_i$ are extreme
states. Necessarily $k>1$ because extreme states cannot have full
rank \cite{agk10}. We deduce that for some $i$ we have
$\bra{\ps}(\r_{A_1B_1}\ox\r_i)^{\G_{A_1A_2}}\ket{\ps}<0$. Therefore
it suffices to use extreme states as the activators in entanglement
distillation.

It is known that the distillable entanglement is upper bounded by
the distillable key \cite{hhh05}. So extreme states can activate the
distillable key of NPT states. Though any PPTES has zero distilalble
entanglement, there are PPTES with positive distillable key
\cite{hhh05}. We may further ask

\textbf{Problem 3}. {\em Can an entangled extreme state produce
distillable key?}


In connection with Theorem \ref{thm:PrvaFam} we raise the following problem.

\textbf{Problem 4}. {\em Construct good $M\times N$ PPTES of rank
$M+N-2$ when $N\ge M\ge4$.}

We propose one more problem about extreme states.

\textbf{Problem 5}. {\em  If $\r$ is a strongly extreme state, is
$\r^\G$ also strongly extreme?}

\acknowledgments

We thank David McKinnon for providing the proof of Proposition
\ref{prop:David}, and Mike Roth for answering our question posed on
MathOverflow \cite{MRoth11} and his help in proving Theorem
\ref{thm:PresekProjVar}. The first author was mainly supported by MITACS
and NSERC. The CQT is funded by the Singapore MoE and the NRF as
part of the Research Centres of Excellence programme. The second
author was supported in part by an NSERC Discovery Grant.

%

\section{Appendix}

Several proofs in the main part of the paper
(Proposition \ref{prop:MxN,>delta=infinitelymany} and
Theorems \ref{thm:delta} and \ref{thm:LMS2010(ii),ker=finitePRO})
rely on two important facts related to the B\'{e}zout's theorem.
Our objective here is to state and prove these facts.

As both proofs make use of the linear projections in projective
space, we shall first sketch their definition.
Let $V\subseteq\bC^{n+1}$ be a vector subspace of
dimension $m+1$ and $L\subseteq\cP^n$ the projective subspace
associated to $V$; its points are the one-dimensional subspaces of
$V$. Let us choose $n-m$ linear forms $l_k:\bC^{n+1}\to\bC$,
$k=1,\ldots,n-m$ such that $V=\cap_k \ker l_k$. Then the map
$\pi:\cP^n\setminus L\to\cP^{n-m-1}$ defined by $\pi(\bC
x)=[l_1(x):\ldots:l_{n-m}(x)]$ is regular, and we refer to it as the
projection with center $L$. It can be described geometrically as
follows. We first fix a subspace $W\subseteq\bC^{n+1}$ of dimension
$n-m$ such that $V\cap W=0$. Our projective space $\cP^{n-m-1}$ will
be the subspace of $\cP^n$ associated to $W$. If
$x\in\bC^{n+1}\setminus V$ then the vector subspace spanned by $V$
and $x$, of dimension $m+1$, meets $W$ in a one-dimensional
subspace, say $\bC y$, and we define $\pi(\bC x)=\bC y$.

The proof of the first proposition is due to David McKinnon
\cite{DMcK11}.
 \bpp \label{prop:David}
Let $X$ be an irreducible projective subvariety of $\cP^n$, of
dimension $k$, and let $L$ be a linear subspace of dimension $m$
(strictly less than $n-k$) such that $L\cap X$ is finite.  Then
there is a linear subspace $M$, containing $L$, whose intersection
with $X$ is again finite, and such that the dimension of $M$ is
exactly $n-k$.
 \epp
 \bpf
Consider the linear projection $\pi:\cP^n\setminus L\to\cP^{n-m-1}$
with center $L$. The set $X^0=X\setminus L$ is open in $X$ and so it
is a quasi-projective variety. Since $\pi$ is a regular map, so is
its restriction $f:X^0\to\cP^{n-m-1}$. The fibres of $f$ are the
intersections of $X^0$ with linear subspaces of dimension $m+1$
containing $L$. Since $m<n-k$, we deduce that $n-m-1\ge k$, so that
$\cP^{n-m-1}$ has dimension at least as large as the dimension of
$X$.

Let $Y$ be the Zariski-closure of the set $f(X^0)$. If $Y$ is not
equal to $\cP^{n-m-1}$, then $f$ is not onto, and so there is some
linear subspace of dimension $m+1$, containing $L$, whose
intersection with $X$ is contained in $L$, and therefore finite. If
$Y$ is equal to $\cP^{n-m-1}$, then there is some nonempty
Zariski-open subset of $Y$ contained in $f(X^0)$ such that the
dimension of the fibre over any of its points plus the dimension of
$Y$ equals the dimension of $X$, see \cite[Corollary (3.15)]{mfd76}.
Consequently, all these fibres have dimension zero, which means that
there exist linear subspaces of dimension $m+1$ containing $L$ whose
intersection with $X$ is finite. In either case, if $m$ is strictly
less than $n-k$, we can construct a linear subspace of dimension
$m+1$ that contains $L$ and still intersects $X$ in a finite set of
points. Continuing in this manner, we can construct the desired
space $M$.
 \epf

The question whether the theorem below is valid was posed on
MathOverflow by the second author
(under additional hypothesis that $X$ is smooth).
The first proof was given by Mike Roth \cite{MRoth11}.
Subsequently, together with Mike, we found another proof given below.
We say that a projective subvariety $X$ of $\cP^n$ is {\em degenerate}
(in $\cP^n$) if it is contained in a hyperplane of $\cP^n$.

 \bt \label{thm:PresekProjVar}
Let $X\subseteq\cP^n$ be an irreducible complex projective variety
embedded in the $n$-dimensional projective space. Let $k$ be the
dimension of $X$ and $d$ its degree. Let $L\subseteq\cP^n$ be a
linear subspace of dimension $n-k$ and $Z=L\cap X$. If $X$ is not
contained in any hyperplane of $\cP^n$ and $Z$ is finite of
cardinality $d$, then $Z$ spans $L$.
 \et
 \bpf
Let $M$ be the linear subspace spanned by $Z$. Assume that $M\ne L$,
and let $m$ $(<n-k)$ be its dimension. We use induction on $m$ to
show that $X$ is degenerate (which contradicts our hypothesis). The
inductive steps will make use of a projection
$\pi:\cP^n\setminus\{p\}\to\cP^{n-1}$ from a suitably chosen point
$p\in M\setminus Z$. We define the maps $f:X\to Y:=\pi(X)$ and
$g:X\to\cP^{n-1}$ to be the restrictions of $\pi$. Since $L$ and $X$
intersect transversely at any $z\in Z$, the differential of $g$ at
$z$ will be injective. Hence, there will exist an open connected
neighborhood $W_z\subseteq X$ of $z$ in analytic (i.e., ordinary)
topology such that $f(W_z)=g(W_z)$ is a complex submanifold of
$\cP^{n-1}$ of dimension $k$ and $f$ induces an isomorphism of $W_z$
and $f(W_z)$ as complex manifolds.

First, suppose that $m=1$, i.e., $M$ is a projective line. Then we
can choose for $p$ any point in $M\setminus Z$. Let $z\in Z$ and
observe that the fibre of $f$ over the point $y_0=f(z)$ is $Z$.

We claim that $Y$ is a cone with vertex $y_0$. Let $y$ be any other
point of $Y$ and $\ell$ the line in $\cP^{n-1}$ joining $y_0$ and
$y$. Suppose that $Y\cap\ell$ is a finite set, and let $P$ be the
2-plane $\{p\}\cup\pi^{-1}(\ell)$. Since each fibre of $f$ is
finite, $P\cap X$ is a finite set. As $P\supset Z\cup f^{-1}(y)$ and
$f^{-1}(y)\ne\emptyset$, we have $|P\cap X|\ge d+1$. By Proposition
\ref{prop:David} there is a $(n-k)$-plane $Q$ such that $Q\supseteq
P$ and $Q\cap X$ is finite. This contradicts the B\'{e}zout's theorem
because $|Q\cap X|>d$. We conclude that the set $Y\cap\ell$ must be
infinite, and so $\ell\subseteq Y$ and our claim is proved.

We next claim that $Y$ is locally irreducible near $y_0$ in the
analytic topology. Suppose on the contrary that $Y$ is locally
reducible near $y_0$, and let $h_1$,\ldots, $h_s$ be local analytic
equations near $y_0$ cutting out an analytic component $V$. The fact
that $Y$ is a cone with vertex $y_0$ then implies (by observing that
the cone remains invariant under scaling) that the homogeneous
pieces of each $h_i$ vanish on $V$, and hence cut out $V$. Since the
homogeneous pieces are homogeneous polynomials, this now implies
that $Y$ is reducible in the Zariski topology. Since $Y$ is
irreducible this is a contradiction and establishes the second
claim.

From the preliminary remarks made above it follows that there exists
an open connected neighborhood $U\subseteq\cP^{n-1}$ of $y_0$ in
analytic topology such that $U\cap Y$ is a union of $d$ complex
$k$-dimensional submanifolds (one for each point  $z\in Z$) passing
through $y_0$. Since the local analytic structure of $Y$ near $y_0$
is a union of $d$ submanifolds, the only way that $Y$ can be
irreducible near $y_0$ in the analytic topology is if all the
submanifolds are the same, so that $Y$ is smooth at $y_0$. This
implies that $Y$ is a linear space of dimension $k$. Hence the
Zariski closure of $f^{-1}(Y)$ is a linear space of dimension $k+1$.
As this linear space contains $X$ and $k+1<n$ (since $1=m < n-k$),
$X$ is degenerate.

Next, suppose that $m>1$. In this case we choose for $p$ a point in
$M\setminus Z$ which is not on any line joining two points of $Z$.
Observe that, for $z\in Z$, the fibre of $f$ over $f(z)$ is $\{z\}$.
Let $W_z$ be chosen as in the beginning of the proof. Since the set
$X'=X\setminus W_z$ is compact in analytic topology, its image
$f(X')$ is also compact. Hence, the set $U=Y\setminus f(X')$ is open
in $Y$ in analytic topology. Note that $f(z)\in U$ and that for each
$y\in U$ there is a unique $x\in W_z$ such that $f(x)=y$ and
$f^{-1}(y)=\{x\}$. On the other hand there exists a nonempty Zariski
open subset $V$ of $Y$ such that for all $y\in V$ the fibre
$f^{-1}(y)$ consists of exactly $d'$ points, where $d'$ is the
degree of $f$. Since $V$ is dense in $Y$, not only in Zariski but
also in analytic topology \cite[Theorem (2.33)]{mfd76}, we conclude
that $U\cap V\ne\emptyset$. Consequently, we have $d'=1$ and the
image $Y$ must have degree $d$ in $\cP^{n-1}$.

The image $f(L\setminus\{p\})$ is a linear subspace of $\cP^{n-1}$
of dimension $(n-1)-k$. Its intersection with $Y$ is the set $f(Z)$
of cardinality $d$. The span of $f(Z)$ is the linear space
$f(M\setminus\{p\})$ of dimension $m-1$. By the induction hypothesis
$Y$ is degenerate in $\cP^{n-1}$, and so $X$ is degenerate in
$\cP^n$.

This completes the proof that $X$ is degenerate, and we can conclude
that our assumption is false, i.e., we must have $M=L$.
 \epf

\end{document}